\documentclass{article}
\usepackage{jheppub} 
\usepackage[utf8x]{inputenc}	
\usepackage[english]{babel}

\usepackage[ddmmyyyy,hhmmss]{datetime}

\usepackage{multirow}

\catcode`@=11
\allowdisplaybreaks
\usepackage{amsxtra}
\usepackage{mathtools}
\usepackage{float}
\usepackage{xcolor}
\usepackage{tikz}
\usepackage{booktabs}
\usepackage{macros}
\usepackage{hyperref}
\usepackage{amsmath}
\usepackage{graphicx}
\usepackage{rotating}  

\usepackage{placeins}
\usetikzlibrary{calc}
\usetikzlibrary{arrows.meta, bending, patterns}
\usepackage{soul}
\usetikzlibrary{hobby,intersections} 
\usetikzlibrary{shapes.geometric}
\usetikzlibrary{shapes.misc}

\tikzset{flavour/.style={draw=none,minimum size=0.3mm,fill=white, regular polygon,regular polygon sides=4,draw}}
\tikzset{flavourr/.style={draw=none,minimum size=0.3mm,fill=red, regular polygon,regular polygon sides=4,draw}}
\tikzset{flavourb/.style={draw=none,minimum size=0.3mm,fill=blue, regular polygon,regular polygon sides=4,draw}}
\tikzset{flavourblack/.style={draw=none,minimum size=0.3mm,fill=black, regular polygon,regular polygon sides=4,draw}}
\tikzset{gaugeBig/.style={inner sep=1mm,draw=none,fill=white,minimum size=2mm,circle, draw}}
\tikzset{bd/.style={circle, draw=black, inner sep=0pt, fill=black, minimum size=2mm}}
\tikzset{wd/.style={circle, draw=black, inner sep=0pt, fill=white, minimum size=2mm}}
\tikzset{Dynkin/.style={circle, draw=black, inner sep=0pt, fill=white, minimum size=2mm}}
\tikzstyle{ligne}=[draw, very thick] 
\tikzstyle{gridline}=[draw, gray] 
\tikzset{gauge/.style={circle, draw,inner sep=2.5pt}}
\tikzset{gaugeo/.style={circle, draw,inner sep=2.5pt,fill=orange}}
\tikzset{gauger/.style={circle, draw,inner sep=2.5pt,fill=red}}
\tikzset{gaugeb/.style={circle, draw,inner sep=2.5pt,fill=blue}}
\tikzset{gaugeg/.style={circle, draw,inner sep=2.5pt,fill=teal}}
\tikzset{gaugegoodteal/.style={circle, draw,inner sep=2.5pt,fill=goodteal}}
\tikzset{gaugem/.style={circle, draw,inner sep=2.5pt,fill=magenta}}
\tikzset{hasse/.style={circle, fill,inner sep=2pt}}
\tikzset{d2/.style={circle, fill,inner sep=1.3pt}}
\tikzset{shrinky/.style={circle, fill,inner sep=1pt}}
\tikzset{sized/.style={circle, draw, inner sep=1.5pt}}
\tikzset{seven/.style={circle, draw,inner sep=3pt}}

\tikzset{gaugebl/.style={circle,draw=black,fill=black,inner sep=1.5pt}}
\tikzset{gaugeblnormal/.style={circle,draw=black,fill=black,inner sep=2.5pt}}
\tikzset{hasse/.style={circle, fill,inner sep=2pt}}
\tikzstyle{dashed_brane}=[thick, dashed]
\tikzstyle{dotted_brane}=[thick, dotted]
\tikzstyle{O3plus}=[thick, color=teal]
\tikzstyle{O3minustilde}=[thick, color=blue]
\tikzstyle{O3plustilde}=[thick, color=red]
\tikzset{D5/.style={cross out, draw=black, minimum size=7, inner sep=0pt, outer sep=0pt}, cross/.default={1pt}}
\tikzset{flavor/.style={regular polygon,regular polygon sides=4,inner sep=2.5pt, label = {}, draw}}
\tikzset{redflavor/.style={regular polygon,regular polygon sides=4,inner sep=2.5pt, color=red, label = {}, draw}}
\tikzset{redgauge/.style={inner sep=1mm,color=red,draw=none,minimum size=2mm,circle, draw}}
\tikzset{blueflavor/.style={regular polygon,regular polygon sides=4,inner sep=2.5pt, color=blue, label = {}, draw}}
\tikzset{bluegauge/.style={inner sep=1mm,color=blue,draw=none,minimum size=2mm,circle, draw}}

\makeatletter
\DeclareRobustCommand{\rvdots}{%
  \vbox{
    \baselineskip4\p@\lineskiplimit\z@
    \kern-\p@
    \hbox{.}\hbox{.}\hbox{.}
  }}
\makeatother

\newcommand{\Figref}[1]{Figure~\ref{#1}}
\newcommand{\Quiver}[1]{$\mathcal Q_{\ref{#1}}$}
\newcommand{\surm}{\mathrm{SU}}

\newcommand{\urm}{\mathrm{U}}
\newcommand{\sorm}{\mathrm{SO}}
\newcommand{\orm}{\mathrm{O}}
\newcommand{\sprm}{\mathrm{Sp}}
\newcommand{\hs}{\mathrm{HS}}

\newcommand{\hwg}{\mathrm{HWG}}
\newcommand{\pe}{\mathrm{PE}}

\newcommand{\hsC}[1]{\hs\left[\mathcal C\left(\text{\Quiver{#1}}\right)\right]}
\newcommand{\hsH}[1]{\hs\left[\mathcal H\left(\text{\Quiver{#1}}\right)\right]}

\tikzset{gaugebl/.style={circle,draw=black,fill=black,inner sep=1.5pt}}
\tikzset{gaugeblnormal/.style={circle,draw=black,fill=black,inner sep=2.5pt}}
\tikzset{hasse/.style={circle, fill,inner sep=2pt}}
\tikzstyle{dashed_brane}=[thick, dashed]
\tikzstyle{dotted_brane}=[thick, dotted]
\tikzstyle{O3plus}=[thick, color=teal]
\tikzstyle{O3minustilde}=[thick, color=blue]
\tikzstyle{O3plustilde}=[thick, color=red]
\tikzset{D5/.style={cross out, draw=black, minimum size=9, inner sep=0pt, outer sep=0pt}, cross/.default={1pt}}
\tikzset{O5circle/.style={circle, minimum size=4mm,draw=black}}
\tikzset{flavor/.style={regular polygon,regular polygon sides=4,inner sep=2.5pt, label = {}, draw}}
\tikzset{redflavor/.style={regular polygon,regular polygon sides=4,inner sep=2.5pt, color=red, label = {}, draw}}
\tikzset{redgauge/.style={inner sep=1mm,color=red,draw=none,minimum size=2mm,circle, draw}}
\tikzset{blueflavor/.style={regular polygon,regular polygon sides=4,inner sep=2.5pt, color=blue, label = {}, draw}}
\tikzset{bluegauge/.style={inner sep=1mm,color=blue,draw=none,minimum size=2mm,circle, draw}}
\allowdisplaybreaks

\usepackage{todonotes}

\title{Quotient Quiver Subtraction -- Classical Groups}

\author{Sam Bennett,}
\author{Amihay Hanany,}
\author{Guhesh Kumaran}

\affiliation{Abdus Salam Centre for Theoretical Physics, Imperial College London,\\ Prince Consort Road
London, SW7 2AZ, UK}
\emailAdd{samuel.bennett18@imperial.ac.uk}
\emailAdd{amihay.hanany@imperial.ac.uk}
\emailAdd{guhesh.kumaran18@imperial.ac.uk}

\preprint{Imperial/TP/26/AH/02}
\abstract{Quotient quiver subtraction is a simple combinatorial prescription for gauging Coulomb branch isometry subgroups of 3d $\mathcal{N}=4$ quiver gauge theories. This paper uses Type IIB brane constructions with $\mathrm{O5}$ planes to extend the prescription to gauge $\sprm(n),\;\sorm(n)$, and $\sprm(n)$ coupled to a half-hypermultiplet Coulomb branch isometry subgroups of quivers with unitary gauge groups. The gauging procedure is no longer solely a subtraction -- additional steps change the graph type. The method is applied to provide alternative constructions of the Higgs branch of certain SCFTs in higher dimensions.}
\begin{document}
\maketitle
\section{Introduction}
This paper expands on the concept of \textit{quotient quiver subtraction} \cite{Hanany:2023tvn,Bennett:2024llh,Bennett:2025edk}, which translates the question of gauging IR Coulomb branch isometry subgroups of a 3d $\mathcal{N}=4$ quiver into a straightforward procedure. Importantly, the algorithm bypasses complications from strongly coupled and non-perturbative physics that arise in the IR; as such, Coulomb branch gauging becomes, for a large class of theories, as simple to perform as flavour gauging. Gauging global symmetries and identifying obstructions to their gauging, is a classic problem in physics and identifying convenient ways to do this is tied to the heart of studying gauge theories.

In many cases the gauging of a Coulomb branch isometry subgroup of a theory is equivalent to the flavour symmetry gauging of the same subgroup of a 3d mirror theory \cite{Intriligator:1996ex}. Quotient quiver subtraction and other methods for gauging Coulomb branch isometry subgroups \cite{Hanany:2024fqf, Dancer:2024lra} are able to go beyond since there are many moduli spaces which do not admit known descriptions as a Higgs branch hyper-Kähler quotient \cite{Hitchin:1986ea} but instead as a moduli space of dressed monopole operators or also as dressed instanton operators \cite{Hanany:2025ctg,Hanany:2025jwo}. Examples include the moduli space of exceptional group instantons on $\mathbb C^2$ which have descriptions as a Coulomb branch where Higgs branch descriptions are unknown \cite{Cremonesi:2014xha}.

A broader class of these moduli spaces are Higgs branches of theories with eight supercharges in four, five and six dimensions \cite{Cabrera:2019izd,Cabrera:2019dob,Bourget:2020asf,Bourget:2020mez,Bourget:2021csg,Lawrie:2023uiu,Mansi:2023faa,Lawrie:2024zon}. In these theories there may be certain phases where additional massless states arise which may not have a perturbative or even a Lagrangian description. To bypass this, the notion of magnetic quivers provides a description of these Higgs branches as a moduli space of monopole operators. Flavour symmetry gauging in these theories with eight supercharges may be described by quotient quiver subtraction or related operations on the magnetic quiver. This motivates the discovery of combinatorial operations on $3d\;\mathcal N=4$ quivers, such as quotient quiver subtraction, as they may be directly used to study non-trivial physics in other dimensions.

Quotient quiver subtraction algorithms have been used so far to gauge $\surm(n)$ Coulomb branch global symmetry subgroups of unitary quiver theories \cite{Hanany:2023tvn}, $\sorm(n)$ and $\sprm(n)$ subgroups of framed orthosymplectic quivers \cite{Bennett:2025edk}, and $\surm(2),\;\surm(3),\;G_2,\;$and $\sorm(7)$ subgroups of unframed orthosymplectic quivers \cite{Bennett:2024llh}. This list is extended here to gauging $\sprm(n),\;\sorm(n),$ and $\sprm(n)$ coupled to a half-hypermultiplet Coulomb branch global symmetry subgroups, using Type IIB brane systems with $\mathrm{O}5$ planes and their mirror the $\mathrm{ON}$. The resulting algorithms involve changing the graph type to that of Dynkin type $D$ or type $C$ for $\sprm(n)$ and $\sorm(n)$ respectively.

The prescription for quotient quiver subtraction, much like previous quotient quiver subtraction and quiver polymerisation algorithms, applies to quivers with a long enough leg of gauge nodes going from $\urm(1)-\urm(2)-\cdots$ which supports the symmetry being gauged.
Furthermore it deals only with cases with \emph{complete Higgsing} of the symmetry being gauged.

The $\sprm(n)$ and $\sorm(n)$ quotient quiver subtraction are used to provide alternative checks of certain $4d\;\mathcal N=2$ dualities in \cite{Argyres:2007tq}, with a more direct approach to deriving magnetic quivers. Further examples provide new relationships between the Higgs branches of certain rank-1 $4d\;\mathcal N=2$ theories under gauging.

\paragraph{Organisation of the paper.} The Type IIB brane systems and notations used throughout this paper are reviewed in Section \ref{sec:Brane}. The concept of Weyl integration is introduced in Section \ref{sec:Weyl} which will be the main computational check of the quotient quiver subtraction algorithms. Section \ref{sec:Sp(n)QQS} introduces the quotient quiver subtraction process for $\sprm(n)$ groups, Sections \ref{sec:So(2n)QQS} and \ref{sec:So(2n+1)QQS} repeat the process for $\sorm(2n)$ and $\sorm(2n+1)$ groups respectively. The case of gauging an $\sprm(n)$ with the addition of a half hypermultiplet is done in Section \ref{sec:Sp1FQQS}. Future directions are discussed in Section \ref{sec:outlook}.

\section{Type IIB Brane Systems}
\label{sec:Brane}
The Type IIB brane systems used throughout this paper are the Hanany-Witten \cite{Hanany:1996ie} brane systems with $\mathrm{D}3,\;\mathrm{D}5,\;\mathrm{NS}5$ brane systems in the presence of $\mathrm{O5}$ and $\mathrm{ON}$ planes. The spacetime occupation of branes and orientifold planes are summarised in Table \ref{tab:BraneOccupation3d}.
\begin{table}[h!]
    \centering
    \begin{tabular}{cccc|ccc|c|ccc}
    \toprule
         &$x^0$&$x^1$ &$x^2$ &$x^3$ &$x^4$ &$x^5$ &$x^6$ &$x^7$ &$x^8$ &$x^9$ \\
         \midrule
         $\mathrm{NS}5/\mathrm{ON}$& $\checkmark$ &$\checkmark$ &$\checkmark$ &$\checkmark$ &$\checkmark$ &$\checkmark$ & & & &\\
         $\mathrm{D}5/\mathrm{O}5$& $\checkmark$ &$\checkmark$ &$\checkmark$& & && & $\checkmark$ &$\checkmark$ &$\checkmark$\\
         $\mathrm{D}3$& $\checkmark$ &$\checkmark$ &$\checkmark$& & & &$\checkmark$& & & \\\bottomrule
    \end{tabular}
    \caption{Occupation of $\mathrm{NS}5, \mathrm{D}5$, and $\mathrm{D}3$ branes in spacetime for Hanany-Witten setup.}
    \label{tab:BraneOccupation3d}
\end{table}

In terms of the brane diagrams, $\mathrm{D}3$ branes will be denoted by horizontal lines, $\mathrm{D}5$ and $\mathrm{O5}$ will be denoted by crosses and unfilled circles respectively, whilst $\mathrm{NS}5$ and $\mathrm{ON}$ will be denoted by vertical solid lines and dashed lines respectively.

\subsection{Reading quivers from brane systems}
The rules for reading off quivers from brane systems in the presence of $\mathrm{O5}$ or $\mathrm{ON}$ planes follows those established in \cite{Hanany:1996ie, Hanany:1999sj, Feng:2000eq, Hanany:2001iy} but it is worth reviewing them once more. A stack of $n\;\mathrm{D}3$ branes suspended between two $\mathrm{NS}5$ branes gives rise to an electric $\urm(n)$ vector multiplet, in quiver notation this will be denoted by an unfilled circle with label $n$. If there are also $k\;\mathrm{D}5$ branes in this $\mathrm{NS}5$ brane interval, there is a contribution of $k$ bifundamental hypermultiplets with flavour symmetry $\urm(k)$. The flavours are denoted with a box with label $k$ and the bifundamental hypermultiplets are denoted with an edge. These rules receive a slight modification in the presence of $\mathrm{O5}$ or $\mathrm{ON}$ planes.

Firstly when $2n\;\mathrm{D}3$ branes pass through an $\mathrm{O5}^-$ plane, the gauge theory on the $\mathrm{D}3$ worldvolume is an $\sprm(n)$ gauge theory. Similarly, when $2n\;\mathrm{D}3$ branes end on an $\mathrm{O5}^+$ plane, there is an $\sorm(2n)$ gauge theory on the $\mathrm{D}3$ worldvolume. In the presence of an $\widetilde{\mathrm{O5}^+}$, the theory on a stack of $2n+1\;\mathrm{D}3$ is $\sorm(2n+1)$ gauge theory. The last case is when $2n\;\mathrm{D}3$ branes pass through an $\widetilde{\mathrm{O}5^-}$, in which case the resulting gauge theory is $\sprm(n)$ with a half-hypermultiplet.

In the case of $\mathrm{ON}$ planes which are S-dual to the $\mathrm{O5}$ planes of corresponding type, the type of gauge symmetry remains unitary but the graph shape of the quiver changes forming a Dynkin diagram shape. When $n\;\mathrm{D3}$ branes pass through $\mathrm{ON}^-$, one chooses a partition of $n$ into $l$ and $n-l$ (where $n>l$), and the resulting quiver has a bifurcation of $\urm(l)$ and $\urm(n-l)$ gauge nodes. The quiver forms a D-type Dynkin diagram shape. In the presence of $\mathrm{ON}^+$ and $\widetilde{\mathrm{ON}^+}$, $n\;\mathrm{D}3$ branes, where $n$ is even and odd respectively, give rise to a non-simply laced edge of multiplicity two with $\urm(n)$ being the long side. The quiver forms a Dynkin diagram shape of C-type. Finally, when $2n\;\mathrm{D}3$ end on an $\widetilde{\mathrm{ON}^-}$ the quiver forms a Dynkin diagram shape of B-type, with a short $\urm(n)$ gauge node with non-simple lacing of multiplicity two.
\section{Weyl Integration}
\label{sec:Weyl}
The effect of gauging global symmetry subgroups on a moduli space $\mathcal M$ can be seen through its refined Hilbert series by performing Weyl integration.  Suppose $\mathcal M$ has global symmetry $G_{\mathcal M}$ and refined Hilbert series $\hs\left[\mathcal M\right](t;u_1,\cdots,u_{\textrm{rank}(G_{\mathcal M})})$, where $t$ and $u_i$ are are fugacities for conformal dimension and global symmetries respectively. Throughout this paper $\mathcal M$ will be a Coulomb branch (or moduli space of dressed monopole operators) so this Hilbert series may be computed with the monopole formula \cite{Cremonesi:2013lqa} or via Hall-Littlewood methods \cite{Cremonesi:2014kwa,Hanany:2019tji}. Gauging a subgroup $G$, where $G\times G'\subset G_{\mathcal M}$ (with $G'$ being the commutant of $G$ inside $G_{\mathcal M}$),  corresponds to a Weyl integration on $\hs\left[\mathcal M\right]$ as shown in \eqref{eq:weyl}.
\begin{equation}
    \hs\left[\mathcal M///G\right](t;x_1,\cdots,x_{r})=\oint_{G}d\mu_G\frac{ \hs\left[\mathcal M\right](t;x_1,\cdots,x_r,y_1,\cdots,y_{\mathrm{rank}(G)})}{\pe[\chi_{\mathrm{Adj}}^G(y_1,\cdots,y_{\mathrm{rank}(G)})t^2]},
\label{eq:weyl}
\end{equation}
Note that $x_i$ are fugacities for the ungauged symmetry subgroup $G'$ of rank $r$, and the $y_i$ are fugacities for the gauged symmetry subgroup $G$.

The equation \eqref{eq:weyl} has a simple physical interpretation. The plethystic exponential in the denominator gives symmetrisations of the adjoint representation of $G$ graded by $t^2$, which have the interpretation of imposing additional F-terms that arise when $G$ is gauged \cite{Feng:2007ur,Forcella:2008bb}. If $G$ is gauged with complete Higgsing (which is assumed and is the case throughout), the $F^{\flat}$-space is a complete intersection and so the (quaternionic) dimension of $\mathcal{M}$ reduces by $\mathrm{dim}\,G$. This change in dimension hence provides a simple check of complete Higgsing.
\section{\boldmath$\sprm(n)\;$\unboldmath Quotient Quivers}
\label{sec:Sp(n)QQS}
An extension of the unitary Coulomb branch gauging algorithm given in \cite{Hanany:2023tvn} involves classical Lie groups of $BCD$-type. In the following derivation, the logic of appealing to the mirror-theory is made clear, alongside its underpinning in brane systems.
\begin{equation}
\raisebox{-0.5\height}{
\begin{tikzpicture}
\node (a) at (0,0){$\begin{tikzpicture}
    \node[gauge, label=below:$N$] (n)at (0,0){};
    \node[flavour, label=above:$2N+k$] (flav) at (0,1){};
    \draw[-] (n)--(flav);
\end{tikzpicture}$};
\node (b) at (6,0){$\begin{tikzpicture}
    \node[gauge, label=below:$1$] (1l) at (0,0){};
    \node[] (cdotsl) at (1,0){$\cdots$};
    \node[gauge, label=below:$N$] (nl) at (2,0){};
    \node[] (cdotsm) at (3,0){$\cdots$};
    \node[gauge, label=below:$N$] (nmr) at (4,0){};
    \node[] (cdotsr) at (5,0){$\cdots$};
    \node[gauge, label=below:$1$] (1r) at (6,0){};
    \node[flavour, label=above:$1$] (1fl) at (2,1){};
    \node[flavour, label=above:$1$] (1fr) at (4,1){};
    \draw[-] (1l)--(cdotsl)--(nl)--(cdotsm)--(nmr)--(cdotsr)--(1r) (1fl)--(nl) (1fr)--(nmr);
    \draw [decorate, decoration = {brace, raise=15pt, amplitude=5pt}] (4.25,0) --  (1.75,0) node[pos=0.5,below=20pt,black]{$k+1$};
    \end{tikzpicture}$};
\node at (1.5,-0.25) [scale=1]{$\longleftrightarrow$};
\end{tikzpicture}}
\label{quiv:UN_SQCD}
\end{equation}
Consider the $\urm(N)$ SQCD with $2N+k$ flavours in the left hand side of \eqref{quiv:UN_SQCD}. Gauging an $\sprm(n)$ subgroup of the flavour symmetry returns the theory \Quiver{eq:UnSprtheory} in \Figref{eq:UnSprtheory} with Type IIB brane system realisation in \Figref{fig:UnSprBrane}. The 3d mirror theory of \Quiver{eq:UnSprtheory} is \Quiver{eq:UnSprtheoryMirror} drawn in \Figref{eq:UnSprtheoryMirror}. The corresponding S-dual brane system is given in \Figref{fig:UnSprMirrorBrane}.
\begin{figure}[h!]
    \centering
    \begin{subfigure}{0.49\textwidth}
    \centering
    \begin{tikzpicture}
    \node[gauge, label=below:$N$] (n)at (0,0){};
    \node[gaugeb, label=below:$\sprm(n)$] (spr) at (-1,0){};
    \node[flavour, label=above:$2N+k-2n$] (flav) at (0,1){};
    \draw[-] (spr)--(n)--(flav);
    \end{tikzpicture}
    \caption{}
    \label{eq:UnSprtheory}
    \end{subfigure}
    \begin{subfigure}{0.49\textwidth}
    \centering
    \begin{tikzpicture}
    \draw[-] (0,1)--(0,-1) (-2,1)--(-2,-1);
    \node[O5circle, label=below:$\mathrm{O5}^-$] at (-4,0){};
    \draw[-] (0,0)--(-2,0)node[midway, above]{$N$}--(-3.8,0)node[midway, above]{$2n$};
    \node[D5] at (-0.5,0.75){};
    \node[D5] at (-1.5,0.75){};
    \node at (-1,0.75){$\cdots$};
    \draw [decorate, decoration = {brace, raise=10pt, 
         amplitude=5pt}] (-1.5,0.75) --  (-0.5,0.75) node[pos=0.5,above=15pt,black]{$2N+k-2n$};
    \end{tikzpicture}
    \caption{}
    \label{fig:UnSprBrane}
    \end{subfigure}
    \begin{subfigure}{0.9\textwidth}
    \centering
    \begin{tikzpicture}
    \node[gauge, label=below:$1$] (1l) at (0,0){};
    \node[gauge, label=below:$2$] (2l) at (-1,0){};
    \node[] (cdotsl) at (-2,0){$\cdots$};
    \node[gauge, label=below:$N$] (nl) at (-3,0){};
    \node[gauge, label=below:$N$] (nml) at (-4,0){};
    \node[] (cdotsm) at (-5,0){$\cdots$};
    \node[gauge, label=below:$N$] (nmr) at (-6,0){};
    \node[gauge, label=below:$N$] (nr) at (-7,0){};
    \node[gauge, label=below:$N-1$] (nm1) at (-8,0){};
    \node[] (cdotsr) at (-9,0){$\cdots$};
    \node[gauge, label=below:$2n+3$] (2rp3) at (-10,0){};
    \node[gauge, label=left:$2n+2$] (2rp2) at (-11,0){};
    \node[gauge, label=left:$n$] (r) at ({-11-cos(45)},{sin(45)}){};
    \node[gauge, label=left:$n+1$] (rp1) at ({-11-cos(45)},{-sin(45)}){};
    \node[flavour, label=above:$1$] (1fl) at (-3,1){};
    \node[flavour, label=above:$1$] (1fr) at (-7,1){};
    \draw[-] (1l)--(2l)--(cdotsl)--(nl)--(nml)--(cdotsm)--(nmr)--(nr)--(nm1)--(cdotsr)--(2rp3)--(2rp2) (2rp2)--(r) (2rp2)--(rp1) (1fl)--(nl) (1fr)--(nr);
    \draw [decorate, decoration = {brace, raise=15pt, 
         amplitude=5pt}] (-3,0) --  (-7,0) node[pos=0.5,below=20pt,black]{$k+1$};
    \end{tikzpicture}
    \caption{}
    \label{eq:UnSprtheoryMirror}
    \end{subfigure}
    \begin{subfigure}{0.9\textwidth}
    \centering
    \begin{tikzpicture}
        \draw[-] (0,-1)--(0,1) (-1,-1)--(-1,1) (-2,-1)--(-2,1) (-3,-1)--(-3,1) (-4,-1)--(-4,1) (-5,-1)--(-5,1) (-6,-1)--(-6,1) (-7,-1)--(-7,1) (-8,-1)--(-8,1) (-9,-1)--(-9,1) (-10,-1)--(-10,1) (-11,-1)--(-11,1) (-12,-1)--(-12,1);
        \draw[dashed] (-13,-1)--(-13,1)node[pos=1,above]{$\mathrm{ON}^-$};
        \draw[-] (0,0)--(-1,0)node[midway, above]{$1$}--(-2,0)node[midway, above]{$2$};
        \draw[-] (-2,0)--(-2.2,0) (-2.8,0)--(-3,0);
        \node at (-2.5,0){$\cdots$};
        \draw[-] (-3,0)--(-4,0)node[midway, above]{$N$};
        \node[D5] at (-3.5,0.75){};
        \draw[-] (-4,0)--(-5,0)node[midway, above]{$N$};
        \draw[-] (-5,0)--(-5.2,0) (-5.8,0)--(-6,0);
        \node at (-5.5,0){$\cdots$};
        \draw[-] (-6,0)--(-7,0)node[midway,above]{$N$}--(-8,0)node[midway,above]{$N$}--(-9,0)node[midway, above]{$N-1$};
        \node[D5] at (-7.5,0.75){};
        \draw[-] (-9,0)--(-9.2,0) (-9.8,0)--(-10,0);
        \node at (-9.5,0){$\cdots$};
        \draw[-] (-10,0)--(-11,0)node[midway, above]{$2n+2$}--(-12,0)node[midway,above]{$n+1$};
        \draw[-] (-11,0.75)--(-12,0.75)node[midway,above]{$n$}--(-13,0.75)--(-12,0.65);
         \draw [decorate, decoration = {brace, raise=5pt, 
         amplitude=5pt}] (-8,1) --  (-3,1) node[pos=0.5,above=10pt,black]{$k+2$};
    \end{tikzpicture}
    \caption{}
    \label{fig:UnSprMirrorBrane}
    \end{subfigure}
    \caption{Gauging an $\sprm(n)$ flavour subgroup of $\urm(N)$ SQCD with $2N+k$ flavours gives the theory drawn in \Figref{eq:UnSprtheory} with the corresponding Type IIB brane system drawn in the electric phase in \Figref{fig:UnSprBrane}. The 3d mirror theory is drawn in \Quiver{eq:UnSprtheoryMirror} with its corresponding Type IIB brane system in electric phase drawn in \Figref{fig:UnSprMirrorBrane}.}
    \label{fig:enter-label}
\end{figure}
Comparing \Figref{eq:UnSprtheoryMirror} to the quiver in the right hand side of \eqref{quiv:UN_SQCD} leads to a candidate for the quotient quiver, given in \eqref{quiv:spn_quotient_quiver}. \Figref{eq:UnSprtheoryMirror} also shows that a second operation, here termed ``splitting'', is necessary in order to correctly realise the gauging. Morally, this is the effect of an $\mathrm{ON}^{0}$ plane. Note that the total rank of the quotient quiver \eqref{quiv:spn_quotient_quiver} is $n(2n+1)$, equal to the dimension of $\sprm(n)$ -- the splitting operation does not change the rank of the theory. Note also that rebalancing with an additional $\urm(1)$ gauge node is not required for $\sprm(n)$ gauging.
\begin{equation}
    \raisebox{-0.5\height}{\begin{tikzpicture}
    \node[] (A) at (-3,-0.1){$\sprm(n) \; \text{Quotient Quiver}$};
        \node[gauge, label=below:$1$] (1l) at (0,0){};
        \node[gauge, label=below:$2$] (2l) at (1,0){};
        \node[] (cdots) at (2,0){$\cdots$};
        \node[gauge, label=below:$2n-1$] (2rm1) at (3,0){};
        \node[gauge, label=below:$2n$] (2r) at (4,0){};
        \draw[-] (1l)--(2l)--(cdots)--(2rm1)--(2r);
    \end{tikzpicture}}
\label{quiv:spn_quotient_quiver}
\end{equation}
\paragraph{Formal Statement of the Rule}
The $\sprm(n)$ \hyperref[fig:SpnQQS]{quotient quiver subtraction} procedure gauges an $\sprm(n)$ subgroup of the Coulomb branch symmetry of any generic, framed or unframed unitary target quiver. The precise rules are given below.
\begin{enumerate}
    \item Take a target quiver with a long leg of gauge nodes up to at least $\urm(2n)$ and at least one more non-negatively balanced gauge node, denoted $\urm(j)$. The edge in \textcolor{teal}{teal} denotes some generic connection to the rest of the quiver $Q$, the only restriction is that the long leg must correspond to long roots of the Coulomb branch global symmetry algebra.
    \begin{equation}
    \raisebox{-0.5\height}{\begin{tikzpicture}
    \node[gauge, label=below:$1$] (1) at (0,0){};
    \node[gauge, label=below:$2$] (2) at (1,0){};
    \node[] (cdots) (cdots) at (2,0){$\cdots$};
    \node[gauge, label=below:$2n$] (2r) at (3,0){};
    \node[gauge, label=below:$j$] (2rp1) at (4,0){};
    \node[gauge] (Q) at (5,0){$Q$};
    \draw[-] (1)--(2)--(cdots)--(2r)--(2rp1);
    \draw[teal,very thick] (2rp1)--(Q);
    \end{tikzpicture}}   
    \end{equation}
    \item Align the $\sprm(n)$ quotient quiver against the leg of the target quiver and subtract the ranks of the quotient quiver from those of the target. This deletes the gauge nodes from $\urm(1)$ up to $\urm(2n)$ in the maximal chain, leaving only $\urm(j)$. Rebalancing is not required.
    \begin{equation}
    \raisebox{-0.5\height}{\begin{tikzpicture}
    \node[gauge, label=below:$1$] (1) at (0,0){};
    \node[gauge, label=below:$2$] (2) at (1,0){};
    \node[] (cdots) (cdots) at (2,0){$\cdots$};
    \node[gauge, label=below:$2n$] (2r) at (3,0){};
    \node[gauge, label=below:$j$] (2rp1) at (4,0){};
    \node[gauge] (Q) at (5,0){$Q$};
    \draw[-] (1)--(2)--(cdots)--(2r)--(2rp1);
    \draw[teal,very thick] (2rp1)--(Q);
    \node[gauge, label=below:$1$] (1sub) at (0,-1){};
    \node[gauge, label=below:$2$] (2sub) at (1,-1){};
    \node[] (cdots) (cdotssub) at (2,-1){$\cdots$};
    \node[gauge, label=below:$2n$] (2rsub) at (3,-1){};
    \node[] (minus) at (-1,-1) {$-$};
    \draw[-] (1sub)--(2sub)--(cdotssub)--(2rsub);
    \node[gauge, label=below:$j$] (2rp1r) at (4,-2){};
    \node[gauge] (Qr) at (5,-2){$Q$};
    \draw[teal,very thick] (2rp1r)--(Qr);
    \end{tikzpicture}}   
    \end{equation}
    \item Split the $\urm(j)$ into two gauge nodes of $\urm\left(\lceil j/2\rceil\right)$ and $\urm(\lfloor j/2\rfloor)$, which inherit all links and flavours of the $\urm(j)$ node in \textcolor{teal}{teal}.
    \begin{equation}
    \begin{tikzpicture}
    \node (a) at (0,0){\begin{tikzpicture}
    \node[gauge, label=below:$j$] (2rp1) at (0,0){};
    \node[gauge] (Qr) at (1,0){$Q$};
    \draw[teal,very thick] (2rp1)--(Qr); 
    \end{tikzpicture}};
    \node (b) at (5,0){\begin{tikzpicture}
    \node[gauge, label=left:$\lfloor j/2\rfloor$] (r) at ({1-cos(45)},{sin(45)}){};
    \node[gauge, label=left:$\lceil j/2\rceil$] (rp1) at ({1-cos(45)},{-sin(45)}){};
    \node[gauge] (Q) at (1,0){$Q$};
    \draw[teal,very thick] (r)--(Qr)--(rp1); 
    \end{tikzpicture}};
    \draw[->] (a)--(b) node[midway, above]{\textrm{Split}};
    \end{tikzpicture}
    \end{equation}
\end{enumerate}
\paragraph{Comments}
Note that the case of $\sprm(1)$ \hyperref[fig:SpnQQS]{quotient quiver subtraction} precisely matches that of the $\surm(2)$ quotient quiver subtraction \cite{Hanany:2023tvn}. Note also that the $\sprm(n)$ quotient quiver is ``half'' of the $\surm(2n)$ quotient quiver, indicating the $\mathbb Z_2$ which relates the two groups. 

The change in the Higgs branch dimension after performing $\sprm(n)$ \hyperref[fig:SpnQQQ]{quotient quiver subtraction}  is \begin{align}
     \Delta\;\mathrm{dim}\;\mathcal H&=\left[-n^2-(n+1)^2\right]-\left[\sum_{i=1}^{2n}i(i+1)-\sum_{i=1}^{2n+1}i^2\right]\nonumber\\&=n=\textrm{Rk}\left(\sprm(n)\right)
 \end{align}which is precisely the expected increase in Higgs branch dimension.

The $\sprm(n)$ \hyperref[fig:SpnQQS]{quotient quiver subtraction} requires a long leg of gauge nodes of the form $(1)-(2)-\cdots-(2n)-(j)-$ where the $\urm(2n)$ and $\urm(j)$ nodes do not have to be balanced. Such a tail has at least $2n-1$ balanced gauge nodes contributing at least $\surm(2n)$ Coulomb branch global symmetry. The $\sprm(n)$ is embedded inside this $\surm(2n)$ as\begin{equation}
    [1,0,\cdots,0,1]_{\surm(2n)}\rightarrow[2,0,\cdots,0]_{\sprm(n)}+[0,1,0\cdots,0]_{\sprm(n)}
\end{equation}This explains the origin of the $\sprm(n)$ symmetry that appears in the long leg. There may be an overall enhancement of the Coulomb branch global symmetry in the starting quiver, in which case the $\sprm(n)$ symmetry being gauged may admit an embedding directly into the enhanced global symmetry. Nevertheless, the $\sprm(n)$ symmetry should be thought of as sourced by the long leg of gauge nodes.

As mentioned, $\sprm(1)$ \hyperref[fig:SpnQQS]{quotient quiver subtraction} is identical to $\surm(2)$ quotient quiver subtraction, with many examples contained within \cite{Hanany:2023tvn}. Non-trivial examples of the algorithm begin with $\sprm(2)$, which requires a quiver with a maximal chain up to $\urm(4)$ and at least one more non-negatively balanced gauge node.
\subsection{$\overline{min.E_8}///\sprm(2)$}
Consider $\sprm(2)$ \hyperref[fig:SpnQQS]{quotient quiver subtraction} on the affine $E^{(1)}_8$ quiver, shown in two steps in \Figref{fig:E8Sp2QQS}. The resulting theory is \Quiver{fig:E8Sp2QQS}, with unrefined Coulomb branch Hilbert series given as \begin{equation}
    \hsC{fig:E8Sp2QQS}=\frac{\left(\begin{aligned}1 &+ 36 t^2 + 812 t^4 + 12735 t^6 + 
   150112 t^8 + 1375706 t^{10} + 10079822 t^{12} + 60251599 t^{14} \\&+ 
   298785809 t^{16} + 1245721768 t^{18}+ 4415307757 t^{20} + 
   13424410658 t^{22} + 35275096941 t^{24} \\&+ 80598415736 t^{26} + 
   160929609663 t^{28} + 281921275764 t^{30} + 434675562921 t^{32} \\&+ 
   591245656950 t^{34} + 710641440658 t^{36} + 755494093072 t^{38} + \cdots + 
   t^{76}\end{aligned}\right)}{(1 - t^2)^{19} (1 - t^4)^{19}}
\end{equation}which does not illuminate the identity of the moduli space as a symplectic singularity but does identify the Coulomb branch global symmetry as $\sorm(11)$ \cite{Gledhill:2021cbe}. This is further verified with the computation of the Coulomb branch and Higgs branch Hasse diagrams \cite{Cabrera:2018ann, Bourget:2022tmw, Bennett:2024loi, Bourget:2024mgn} in \Figref{fig:E8Sp2Hasse}. From the Coulomb branch Hasse diagram \Figref{fig:E8Sp2CoulHasse} it is clear that the Hasse diagram for the moduli space of 2 $\sorm(8)$ instantons on $\mathbb C^2$ is a subdiagram. The bottom two slices form the Hasse diagram of $\overline{n. min. B_5}$. From the Higgs branch Hasse diagram, in \Figref{fig:E8Sp2HiggsHasse}, the Namikawa-Weyl group is seen to be $\mathbb Z_2^5\rtimes S_5$. Additionally, the diamond at the bottom is the Hasse diagram of $\textrm{Sym}^2(D_4)$.
\begin{figure}[h!]
    \centering
    \begin{subfigure}[t]{0.5\textwidth}
    \begin{tikzpicture}
        \node[gauge, label=below:$1$] (1) at (0,0){};
        \node[gauge, label=below:$2$] (2) at (1,0){};
        \node[gauge, label=below:$3$] (3) at (2,0){};
        \node[gauge, label=below:$4$] (4) at (3,0){};
        \node[gauge, label=below:$5$] (5) at (4,0){};
        \node[gauge, label=below:$6$] (6) at (5,0){};
        \node[gauge, label=below:$4$] (4r) at (6,0){};
        \node[gauge, label=below:$2$] (2r) at (7,0){};
        \node[gauge, label=right:$3$] (3t) at (5,1){};

        \draw[-] (1)--(2)--(3)--(4)--(5)--(6)--(4r)--(2r) (6)--(3t);

         \node[gauge, label=below:$1$] (1sub) at (0,-1){};
        \node[gauge, label=below:$2$] (2sub) at (1,-1){};
        \node[gauge, label=below:$3$] (3sub) at (2,-1){};
        \node[gauge, label=below:$4$] (4sub) at (3,-1){};
        \node[] (minus) at (-1,-1) {$-$};
        \draw[-] (1sub)--(2sub)--(3sub)--(4sub);

        \node[gauge, label=below:$5$] (5r) at (4,-2){};
        \node[gauge, label=below:$6$] (6r) at (5,-2){};
        \node[gauge, label=below:$4$] (4r) at (6,-2){};
        \node[gauge, label=below:$2$] (2r) at (7,-2){};
        \node[gauge, label=right:$3$] (3r) at (5,-1){};
        \draw[-] (5r)--(6r)--(4r)--(2r) (6r)--(3r);
    \end{tikzpicture}     
    \caption{}
    \label{fig:E8Sp2Sub}
    \end{subfigure}
    \begin{subfigure}[t]{0.5\textwidth}
        \begin{tikzpicture}
            \node (a) at (0,0){\begin{tikzpicture}
                \node[gauge, label=below:$5$] (5r) at (4,-2){};
        \node[gauge, label=below:$6$] (6r) at (5,-2){};
        \node[gauge, label=below:$4$] (4r) at (6,-2){};
        \node[gauge, label=below:$2$] (2r) at (7,-2){};
        \node[gauge, label=right:$3$] (3r) at (5,-1){};
        \draw[-] (5r)--(6r)--(4r)--(2r) (6r)--(3r);
            \end{tikzpicture}};
        \node (b) at (5,0){\begin{tikzpicture}
        \node[gauge, label=below:$2$] (5r) at (4,-2){};
        \node[gauge, label=below:$6$] (6r) at (5,-2){};
        \node[gauge, label=below:$4$] (4r) at (6,-2){};
        \node[gauge, label=below:$2$] (2r) at (7,-2){};
        \node[gauge, label=right:$3$] (3r) at ({5+cos(60)},{-2+sin(60)}){};
        
        \node[gauge, label=left:$3$] (3r2) at ({5-cos(60)},{-2+sin(60)}){};
        \draw[-] (5r)--(6r)--(4r)--(2r) (3r2)--(6r)--(3r);    \end{tikzpicture}};
        \draw[->] (a)--(b) node[midway, above]{\textrm{Split}} node[midway, below]{$\urm(5)$};
        \end{tikzpicture}
    \caption{}
    \label{fig:E8Sp2Split}
    \end{subfigure}
    \caption{$\sprm(2)$ \hyperref[fig:SpnQQS]{quotient quiver subtraction} on the affine $E^{(1)}_8$ quiver which occurs in two steps. The first step is shown in \ref{fig:E8Sp2Sub} and involves the subtraction of the $\sprm(2)$ quotient quiver from the maximal leg. The second step, shown \ref{fig:E8Sp2Split}, is the splitting of the $\urm(5)$ node into a $\urm(2)$ and $\urm(3)$, producing the resulting quiver \Quiver{fig:E8Sp2QQS}. Note the emergence of an $S_2$ outer automorphism between the two $\urm(3)$ gauge nodes.}
    \label{fig:E8Sp2QQS}
\end{figure}

\begin{figure}[h!]
    \centering
    \begin{subfigure}[t]{0.45\textwidth}
    \centering\begin{tikzpicture}
    \node[hasse] (1) at (0,0){};
    \node[hasse] (2) at (0,-1.5){};
    \node[hasse] (3) at (0,-3){};
    \node[hasse] (4) at (0,-4.5){};
    \node[hasse] (5) at (0,-6){};

    \draw[-] (1)--(2)node[midway, right]{$d_4$}--(3) node[midway, right]{$2d_4$}--(4)node[midway, right]{$A_1$}--(5)node[midway, right]{$b_5$};
    \end{tikzpicture}     
    \caption{}
    \label{fig:E8Sp2CoulHasse}
    \end{subfigure}
    \begin{subfigure}[t]{0.45\textwidth}
    \centering
        \begin{tikzpicture}
        \node[hasse] (1) at (0,0){};
        \node[hasse] (2l) at (-2,2){};
        \node[hasse] (2r) at (2,2){};
        \node[hasse] (3) at (0,4){};
        \node[hasse] (4) at (0,6){};

        \draw[-] (1)--(2l)node[midway, left]{$D_4$}--(3)node[midway, left]{$D_4$}--(4)node[midway, right]{$C_5$} (1)--(2r)node[midway, right]{$D_4$}--(3)node[midway, right]{$A_1$};
            
        \end{tikzpicture}
    \caption{}
    \label{fig:E8Sp2HiggsHasse}
    \end{subfigure}
    \caption{Hasse diagrams for the Coulomb and Higgs branches of \Quiver{fig:E8Sp2QQS} in \Figref{fig:E8Sp2CoulHasse} and \Figref{fig:E8Sp2HiggsHasse} respectively. The Coulomb branch undergoes a symmetry enhancement to $\sorm(11)$, as predicted by \cite{Gledhill:2021cbe}.}
    \label{fig:E8Sp2Hasse}
\end{figure}
The explicit Hilbert series of the hyper-Kähler quotient may be computed with Weyl integration with the following embedding $E_8\hookleftarrow\sorm(11)\times\sprm(2)$ which decomposes the adjoint as \begin{equation}
    \left(\mu_7\right)_{E_8}\rightarrow \mu_2+\nu_1^2+\mu_1\nu_2+\mu_5\nu_1
\end{equation}where the $\mu_{1,2,3,4,5}$ and $\nu_{1,2}$ are highest weight fugacities for $\sorm(11)$ and $\sprm(2)$ respectively.

The result from Weyl integration confirms that \begin{equation}
    \overline{min. E_8}///\sprm(2)=\mathcal C\left(\mathcal Q_{\ref{fig:E8Sp2QQS}}\right)
\end{equation}

This construction is a realisation of the proposed duality of certain $4d\;\mathcal N=2$ theories in \cite[Table 2 Ex. 1]{Argyres:2007tq} through magnetic quivers. The pertinent example here is that of the $\sprm(2)$ flavour symmetry gauging of the rank 1 $E_8$ SCFT \cite{Dasgupta:1996ij,Morrison:1996na, Morrison:1996pp, Minahan:1996cj} being dual to the following quiver \Quiver{eq:E8Sp2Mirror}\begin{equation}
    \begin{tikzpicture}
        \node[gaugeb, label=below:$\sprm(3)$] (sp3) at (0,0){};
        \node[flavourr, label=above:$\sorm(11)$] (so11) at (0,1){};
        \node[flavourr, label=below:$\orm(1)$] (o1) at (1,0){};

        \draw[-] (sp3)--(so11);
        \draw[-, red] (sp3)--(o1);
    \end{tikzpicture}\label{eq:E8Sp2Mirror}
\end{equation}where the red edge denotes a ``bi-spinor''. In \cite[Table 2]{Argyres:2007tq} this theory is given as $\sprm(3)$ with $\textbf{14}\oplus11\cdot\textbf{6}$. The $11\cdot\textbf{6}$ unambiguously refers to the 11 half-hypers. However, there are two 14-dimensional representations of $\sprm(3)$, the anti-symmetric, $[0,1,0]$, and ``spinor'', $[0,0,1]$ (with the condition that the highest anti-symmetric representation of $\sprm(n)$ is called the spinor). Cancellation of the Witten anomaly \cite{Witten:1982fp} requires the total Dynkin index of the matter representations to be an integer. The eleven half-hypers each contribute $T([1,0,0])=1/2$, the Dynkin index of the spinor is $T([0,0,1])=5/2$ which gives the desired cancellation, whereas the Dynkin index for the anti-symmetric is $T([0,1,0])=2$ which does not give the required cancellation. Furthermore, the continuous commutant of $\sprm(2)$ inside $E_8$ is $\sorm(11)$. The spinor is a pseudoreal representation meaning that there is an additional discrete $\orm(1)$ flavour symmetry. The anti-symmetric representation is real meaning that there would be an additional $\sprm(1)$ flavour symmetry which does not fit with the embedding of $\sprm(2)$ inside $E_8$.

For the purposes of this work, where the focus lies on magnetic quivers, the duality proposed in \cite{Argyres:2007tq} suggests that the Coulomb branch of \Quiver{fig:E8Sp2QQS} and the Higgs branch of \Quiver{eq:E8Sp2Mirror} are the same.

It is straightforward to compute the Higgs branch Hilbert series of \Quiver{eq:E8Sp2Mirror} to find agreement\begin{equation}
    \mathcal H\left(\text{\Quiver{eq:E8Sp2Mirror}}\right)=\mathcal C\left(\text{\Quiver{fig:E8Sp2QQS}}\right)
\end{equation}

In particular this makes \Quiver{fig:E8Sp2QQS} an example of a magnetic quiver for a rank-3 four dimensional theory. Magnetic quivers for rank-1 and rank-2 theories were studied in \cite{Bourget:2020asf, Bourget:2021csg}. The $\orm(1)\simeq\mathbb Z_2$ flavour symmetry in \Quiver{eq:E8Sp2Mirror} is expressed in \Quiver{fig:E8Sp2QQS} as the $\mathbb Z_2$ exchanging the two $\urm(3)$ gauge nodes.


\subsection{$(\overline{min. E_6}\times\mathbb H^{12})///\sprm(2)$}
There is another proposed duality in \cite[Table 2 Ex. 16]{Argyres:2007tq} which states that the $\sprm(2)$ flavour symmetry gauging of the rank 1 $E_6$ SCFT \cite{Dasgupta:1996ij,Morrison:1996na, Morrison:1996pp, Minahan:1996cj} coupled to 12 free hypers is dual to $\surm(4)$ with four fundamental flavours and two anti-symmetric.

The electric quiver is given below \begin{equation}
    \begin{tikzpicture}
        \node[gaugeblack, label=below:$\surm(4)$] (su4) at (0,0){};
        \node[flavour, label=above:$\urm(4)$] (u4) at (0,1){};
        \node[flavourb, label=below:$\sprm(2)$] (sp2) at (1,0){};

        \draw[-] (u4)--(su4);
        \draw[-,color=red](su4)--(sp2) node[midway, above, color=black]{$\wedge^2$};
    \end{tikzpicture}\label{eq:E6H12Sp2Mirror}
\end{equation}with the (unrefined) Higgs branch Hilbert series computed as \begin{equation}
    \hsH{eq:E6H12Sp2Mirror}=\frac{\left(\begin{aligned}
        1 &+ 17 t^2 + 36 t^3 + 220 t^4 + 564 t^5 + 2192 t^6 + 5456 t^7 + 
 15897 t^8 + 35388 t^9 + 84385 t^{10} \\&+ 165040 t^{11} + 330357 t^{12} + 
 558676 t^{13} + 948191 t^{14} + 1360392 t^{15} + 1931582 t^{16} \\&+ 
 2222620 t^{17} + 2460927 t^{18} + 1755844 t^{19} + 745362 t^{20} - 
 1855388 t^{21} - 4328851 t^{22} \\&- 8203044 t^{23} - 10166289 t^{24} - 
 12396256 t^{25} - 10389664 t^{26} - 8125224 t^{27} - 936407 t^{28} \\&+ 
 4603732 t^{29} + 13037971 t^{30} + 16246308 t^{31} + 19779930 t^{32}+\cdots+t^{64}
    \end{aligned}\right)}{(1-t^2)^9(1-t^3)^{12}(1-t^4)^9}
\end{equation} The Hilbert series shows that the moduli space has global symmetry $\urm(4)\times \sprm(2)$ as expected. The (quaternionic) dimension of the Higgs branch is 13 from counting $\textrm{dim}\;\mathcal H\left(\text{\Quiver{eq:E6H12Sp2Mirror}}\right)=\#\text{hypers}-\#\text{vectors}$. This is not immediately apparent in the Hilbert series above due to negative terms in the numerator, however the order of the pole of the Hilbert series at $t=1$ is indeed 26 which means that the quaternionic dimension is 13.

The proposed dual to \Quiver{eq:E6H12Sp2Mirror} takes the rank 1 $E_6$ theory coupled to 12 free hypers and gauges $\sprm(2)$. In terms of magnetic quivers, the rank 1 $E_6$ theory has a magnetic quiver given by the affine $E_6^{(1)}$ quiver. The coupling of twelve free hypers can be done by the method of \textit{quiver extension} \cite{Hanany:2024fqf} which is a form of quiver addition. The quiver extension needed in this example adds the finite $A_4$ quiver three times to the $E_6^{(1)}$ quiver resulting in the following magnetic quiver for the rank 1 $E_6$ theory coupled to 12 free hypers \begin{equation}
    \begin{tikzpicture}
        \node[gauge, label=below:$1$] (1l) at (0,0){};
        \node[gauge, label=below:$2$] (2l) at (1,0){};
        \node[gauge, label=below:$3$] (3l) at (2,0){};
        \node[gauge, label=below:$4$] (4l) at (3,0){};
        \node[gauge, label=below:$4$] (4m) at (4,0){};
        \node[gauge, label=below:$4$] (4r) at (5,0){};
        \node[gauge, label=below:$2$] (2r) at (6,0){};
        \node[gauge, label=below:$1$] (1r) at (7,0){};
        \node[gauge, label=right:$2$] (2t) at (5,1){};
        \node[gauge, label=right:$1$] (1t) at (5,2){};

        \draw[-] (1l)--(2l)--(3l)--(4l)--(4m)--(4r)--(2r)--(1r) (1t)--(2t)--(4r);
    \end{tikzpicture}\label{eq:E6H12Extended}
\end{equation}

The $\sprm(2)$ gauging proceeds with quotient quiver subtraction shown in \Figref{fig:E6H12Sp2QQS} to produce \Quiver{fig:E6H12Sp2QQS}. Computation of the Coulomb branch Hilbert series indeed confirms that \begin{equation}
    \mathcal C\left(\text{\Quiver{fig:E6H12Sp2QQS}}\right)=\mathcal H\left(\text{\Quiver{eq:E6H12Sp2Mirror}}\right)
\end{equation}so \Quiver{fig:E6H12Sp2QQS} is a magnetic quiver for the Higgs branch of the $4d\;\mathcal N=2$ theory \Quiver{eq:E6H12Sp2Mirror}. The Coulomb and Higgs branch Hasse diagrams are shown in \Figref{fig:E6H12Sp2Hasse}.

This magnetic quiver was proposed in \cite{Chacaltana:2010ks} through methods studying the theory \Quiver{eq:E6H12Sp2Mirror} as a class $\mathcal S$ theory and considering its S-dual frames. It may also be found from the explicit Type IIA brane system for $\surm(4)$ gauge theory with four flavours and two anti-symmetric \cite{Hanany:1999sj}. Quotient quiver subtraction provides an alternative construction for the magnetic quiver.

\begin{figure}[H]
    \centering
    \begin{subfigure}[t]{0.5\textwidth}
    \begin{tikzpicture}
        \node[gauge, label=below:$1$] (1l) at (0,0){};
        \node[gauge, label=below:$2$] (2l) at (1,0){};
        \node[gauge, label=below:$3$] (3l) at (2,0){};
        \node[gauge, label=below:$4$] (4l) at (3,0){};
        \node[gauge, label=below:$4$] (4m) at (4,0){};
        \node[gauge, label=below:$4$] (4r) at (5,0){};
        \node[gauge, label=below:$2$] (2r) at (6,0){};
        \node[gauge, label=below:$1$] (1r) at (7,0){};
        \node[gauge, label=right:$2$] (2t) at (5,1){};
        \node[gauge, label=right:$1$] (1t) at (5,2){};

        \draw[-] (1l)--(2l)--(3l)--(4l)--(4m)--(4r)--(2r)--(1r) (1t)--(2t)--(4r);

         \node[gauge, label=below:$1$] (1sub) at (0,-1){};
        \node[gauge, label=below:$2$] (2sub) at (1,-1){};
        \node[gauge, label=below:$3$] (3sub) at (2,-1){};
        \node[gauge, label=below:$4$] (4sub) at (3,-1){};
        \node[] (minus) at (-1,-1) {$-$};
        \draw[-] (1sub)--(2sub)--(3sub)--(4sub);

        \node[gauge, label=below:$4$] (4mres) at (4,-3){};
        \node[gauge, label=below:$4$] (4rres) at (5,-3){};
        \node[gauge, label=below:$2$] (2rres) at (6,-3){};
        \node[gauge, label=below:$1$] (1rres) at (7,-3){};
        \node[gauge, label=right:$2$] (2tres) at (5,-2){};
        \node[gauge, label=right:$1$] (1tres) at (5,-1){};

        \draw[-] (4mres)--(4rres)--(2rres)--(1rres) (1tres)--(2tres)--(4rres);
    \end{tikzpicture}     
    \caption{}
    \label{fig:E6H12Sp2Sub}
    \end{subfigure}
    \begin{subfigure}[t]{0.5\textwidth}
        \begin{tikzpicture}
            \node (a) at (0,0){\begin{tikzpicture}
                \node[gauge, label=below:$4$] (4mres) at (4,-3){};
        \node[gauge, label=below:$4$] (4rres) at (5,-3){};
        \node[gauge, label=below:$2$] (2rres) at (6,-3){};
        \node[gauge, label=below:$1$] (1rres) at (7,-3){};
        \node[gauge, label=right:$2$] (2tres) at (5,-2){};
        \node[gauge, label=right:$1$] (1tres) at (5,-1){};

        \draw[-] (4mres)--(4rres)--(2rres)--(1rres) (1tres)--(2tres)--(4rres);
            \end{tikzpicture}};
        \node (b) at (5,0){\begin{tikzpicture}
        \node[gauge, label=below:$2$] (2mres) at (4,-3){};
        \node[gauge, label=left:$2$] (2split) at ({5-sin(45)},{-3+cos(45)}){};
        \node[gauge, label=below:$4$] (4rres) at (5,-3){};
        \node[gauge, label=below:$2$] (2rres) at (6,-3){};
        \node[gauge, label=below:$1$] (1rres) at (7,-3){};
        \node[gauge, label=right:$2$] (2tres) at (5,-2){};
        \node[gauge, label=right:$1$] (1tres) at (5,-1){};

        \draw[-] (2split)--(4rres) (2mres)--(4rres)--(2rres)--(1rres) (1tres)--(2tres)--(4rres); \end{tikzpicture}};
        \draw[->] (a)--(b) node[midway, above]{\textrm{Split}} node[midway, below]{$\urm(4)$};
        \end{tikzpicture}
    \caption{}
    \label{fig:E6H12Sp2Split}
    \end{subfigure}
    \caption{$\sprm(2)$ \hyperref[fig:SpnQQS]{quotient quiver subtraction} on the extended affine $E^{(1)}_6$ quiver \Quiver{eq:E6H12Extended} which occurs in two steps. The first step is shown in \ref{fig:E6H12Sp2Sub} and involves the subtraction of the $\sprm(2)$ quotient quiver from the maximal leg. The second step, shown \ref{fig:E6H12Sp2Split}, is the splitting of the $\urm(4)$ node into two $\urm(2)$ nodes producing the resulting quiver \Quiver{fig:E6H12Sp2QQS}.}
    \label{fig:E6H12Sp2QQS}
\end{figure}

\begin{figure}[H]
\centering
\begin{subfigure}{0.49\textwidth}
\centering
    \begin{tikzpicture}
    \node[hasse] (1) at (0,0){};
    \node[hasse] (2) at (0,-1.5){};
    \node[hasse] (3) at (0,-3){};
    \node[hasse] (4) at (0,-4.5){};
    \node[hasse] (5) at (0,-6){};
    \node[hasse] (6) at (2,-3.75){};
    \draw[-] (1)--(2)node[midway, left]{$d_4$}--(3) node[midway, left]{$2d_4$}--(4)node[midway, left]{$A_1$}--(5)node[midway, left]{$c_2$} (2)--(6)node[midway, above]{$a_5$} --(5)node[midway, below]{$a_3$};
    \end{tikzpicture}
    \caption{}
    \label{fig:E6H12Sp2CoulHasse}
\end{subfigure}
\begin{subfigure}{0.49\textwidth}
\centering
    \begin{tikzpicture}
    \node[hasse] (1) at (0,0){};
    \node[hasse] (2) at (0,-1.5){};
    \node[hasse] (3) at (-1.5,-3){};
    \node[hasse] (4) at (1.5,-3){};
    \node[hasse] (5) at (0,-4.5){};
    \draw[-] (1)--(2)node[midway, left]{$C_2$}--(3) node[midway, above]{$D_4$}--(5)node[midway, below]{$D_4$} (2)--(4)node[midway, above]{$c_1$}--(5)node[midway, below]{$D_4$};
    \end{tikzpicture}
    \caption{}
    \label{fig:E6H12Sp2HiggsHasse}
\end{subfigure}
\caption{Hasse diagrams for the Coulomb and Higgs branches of \Quiver{fig:E6H12Sp2QQS} shown in \Figref{fig:E6H12Sp2CoulHasse} and \Figref{fig:E6H12Sp2HiggsHasse} respectively.}
\label{fig:E6H12Sp2Hasse}
\end{figure}

The arguments above can be employed to generalise the duality proposed in \cite[Table 2 Ex. 16]{Argyres:2007tq} to a one-parameter family of dualities. Specifically that \begin{equation}
    \mathcal H_{4d}\left(\begin{tikzpicture}[baseline=-0.5ex]
        \node[gaugeblack, label=below:$\surm(2k)$] (su4) at (0,0){};
        \node[flavour, label=above:$\urm(4)$] (u4) at (0,1){};
        \node[flavourb, label=below:$\sprm(2)$] (sp2) at (1,0){};

        \draw[-] (u4)--(su4);
        \draw[-,color=red](su4)--(sp2) node[midway, above, color=black]{$\wedge^2$};
    \end{tikzpicture}\right)=\mathcal H^\infty_{5d}\left(\begin{tikzpicture}[baseline=-0.5ex]
        \node[gaugeb, label=below:$\sprm(k-1)$] (suk) at (0,0){};
        \node[flavourr, label=above:$\sorm(4k+2)$] (ukf) at (0,1){};

        \draw[-] (suk)--(ukf);
    \end{tikzpicture}\right)\times \mathbb H^{6k}///\sprm(k),\quad k\geq2\label{eq:4d5dduality}
\end{equation}

The magnetic quiver for the $5d$ theory can be extended by three nodes of $\urm(2k)$ to incorporate the $\mathbb H^{6k}$ producing the following quiver 

\begin{equation}
    \begin{tikzpicture}
        \node[gauge, label=below:$1$] (1l) at (0,0){};
        \node[gauge, label=below:$2$] (2l) at (1,0){};
        \node (cdots) at (2,0){$\cdots$};
        \node[gauge, label=below:$2k-1$] (2km1) at (3,0){};
        \node[gauge, label=below:$2k$] (2kl) at (4,0){};
        \node[gauge, label=below:$2k$] (2km) at (5,0){};
        \node[gauge, label=below:$2k$] (2kr) at (6,0){};
        \node[gauge, label=below:$k$] (kr) at (7,0){};
        \node[gauge, label=below:$1$] (1r) at (8,0){};
        \node[gauge, label=right:$k$] (kt) at (6,1){};
        \node[gauge, label=right:$1$] (1t) at (6,2){};

        \draw[-] (1l)--(2l)--(cdots)--(2km1)--(2kl)--(2km)--(2kr)--(kr)--(1r) (2kr)--(kt)--(1t); 
        \end{tikzpicture}\label{eq:SUkH6k}
\end{equation}

Then performing $\sprm(k)$ quotient quiver subtraction on \Quiver{eq:SUkH6k} is shown in \Figref{fig:SUkH6kSpkQQS} to produce \Quiver{fig:SUkH6kSpkQQS} which is precisely what one reads from the Type IIA brane system for the $4d$ quiver in \eqref{eq:4d5dduality}.
\begin{figure}[H]
    \centering
    \begin{subfigure}[t]{0.5\textwidth}
    \begin{tikzpicture}
         \node[gauge, label=below:$1$] (1l) at (0,0){};
        \node[gauge, label=below:$2$] (2l) at (1,0){};
        \node (cdots) at (2,0){$\cdots$};
        \node[gauge, label=below:$2k-1$] (2km1) at (3,0){};
        \node[gauge, label=below:$2k$] (2kl) at (4,0){};
        \node[gauge, label=below:$2k$] (2km) at (5,0){};
        \node[gauge, label=below:$2k$] (2kr) at (6,0){};
        \node[gauge, label=below:$k$] (kr) at (7,0){};
        \node[gauge, label=below:$1$] (1r) at (8,0){};
        \node[gauge, label=right:$k$] (kt) at (6,1){};
        \node[gauge, label=right:$1$] (1t) at (6,2){};

        \draw[-] (1l)--(2l)--(cdots)--(2km1)--(2kl)--(2km)--(2kr)--(kr)--(1r) (2kr)--(kt)--(1t);

         \node[gauge, label=below:$1$] (1sub) at (0,-1){};
        \node[gauge, label=below:$2$] (2sub) at (1,-1){};
        \node (cdotssub) at (2,-1){};
        \node[gauge, label=below:$2k-1$] (2km1sub) at (3,-1){};
        \node[gauge, label=below:$2k$] (2ksub) at (4,-1){};
        \node[] (minus) at (-1,-1) {$-$};
        \draw[-] (1sub)--(2sub)--(cdotssub)--(2km1sub)--(2ksub);

        \node[gauge, label=below:$2k$] (2kmr) at (5,-3){};
        \node[gauge, label=below:$2k$] (2krr) at (6,-3){};
        \node[gauge, label=below:$k$] (krr) at (7,-3){};
        \node[gauge, label=below:$1$] (1rr) at (8,-3){};
        \node[gauge, label=right:$k$] (ktr) at (6,-2){};
        \node[gauge, label=right:$1$] (1tr) at (6,-1){};

        \draw[-] (2kmr)--(2krr)--(krr)--(1rr) (2krr)--(ktr)--(1tr); 

    \end{tikzpicture}     
    \caption{}
    \label{fig:SUkH6kSpkSub}
    \end{subfigure}
    \begin{subfigure}[t]{0.5\textwidth}
        \begin{tikzpicture}
            \node (a) at (0,0){\begin{tikzpicture}
               \node[gauge, label=below:$2k$] (2kmr) at (5,-3){};
        \node[gauge, label=below:$2k$] (2krr) at (6,-3){};
        \node[gauge, label=below:$k$] (krr) at (7,-3){};
        \node[gauge, label=below:$1$] (1rr) at (8,-3){};
        \node[gauge, label=right:$k$] (ktr) at (6,-2){};
        \node[gauge, label=right:$1$] (1tr) at (6,-1){};

        \draw[-] (2kmr)--(2krr)--(krr)--(1rr) (2krr)--(ktr)--(1tr); 
            \end{tikzpicture}};
        \node (b) at (5,0){\begin{tikzpicture}
        \node[gauge, label=below:$k$] (kmr) at (5,-3){};
        \node[gauge, label=left:$k$] (kmr2) at ({6-sin(45)},{-3+cos(45)}){};
        \node[gauge, label=below:$2k$] (2krr) at (6,-3){};
        \node[gauge, label=below:$k$] (krr) at (7,-3){};
        \node[gauge, label=below:$1$] (1rr) at (8,-3){};
        \node[gauge, label=right:$k$] (ktr) at (6,-2){};
        \node[gauge, label=right:$1$] (1tr) at (6,-1){};

        \draw[-] (kmr2)--(2krr) (kmr)--(2krr)--(krr)--(1rr) (2krr)--(ktr)--(1tr);  \end{tikzpicture}};
        \draw[->] (a)--(b) node[midway, above]{\textrm{Split}} node[midway, below]{$\urm(2k)$};
        \end{tikzpicture}
    \caption{}
    \label{fig:SUkH6kSpkQQSplit}
    \end{subfigure}
    \caption{$\sprm(2)$ \hyperref[fig:SpnQQS]{quotient quiver subtraction} on \Quiver{eq:SUkH6k} which occurs in two steps. The first step is shown in \ref{fig:SUkH6kSpkSub} and involves the subtraction of the $\sprm(k)$ quotient quiver from the maximal leg. The second step, shown \ref{fig:SUkH6kSpkQQSplit}, is the splitting of the $\urm(2k)$ node into two $\urm(k)$ nodes producing the resulting quiver \Quiver{fig:SUkH6kSpkQQS}.}
    \label{fig:SUkH6kSpkQQS}
\end{figure}

\section{\boldmath$\sorm(2n)\;$\unboldmath Quotient Quivers}
\label{sec:So(2n)QQS}
\begin{figure}
    \centering
    \begin{subfigure}[t]{0.4\textwidth}
        \begin{tikzpicture}
        \node[gauge, label=below:$N$] (n)at (0,0){};
        \node[gauger, label=below:$\sorm(2n)$] (spr) at (-1,0){};
        \node[flavour, label=above:$2N+k-2n$] (flav) at (0,1){};
        \draw[-] (spr)--(n)--(flav);
    \end{tikzpicture}
    \caption{}
    \label{eq:UnSOtheory}
    \end{subfigure}
    \begin{subfigure}[t]{0.4\textwidth}
        \begin{tikzpicture}
        \draw[-] (0,-1)--(0,1) (-2,-1)--(-2,1);
        \node[O5circle,label=below:$\mathrm{O5}^+$] at (-4,0){};

        \draw[-] (0,0)--(-2,0)node[midway, above]{$N$}--(-3.8,0)node[midway, above]{$2n$};
        \node[D5] at (-0.5,0.75){};
        \node[D5] at (-1.5,0.75){};
        \node at (-1,0.75){$\cdots$};
        \draw [decorate, decoration = {brace, raise=10pt, 
         amplitude=5pt}] (-1.5,0.75) --  (-0.5,0.75) node[pos=0.5,above=15pt,black]{$2N+k-2n$};  
    \end{tikzpicture}
    \caption{}
    \label{fig:UnSOBrane}
    \end{subfigure}
    \begin{subfigure}[t]{0.75\textwidth}
    \begin{tikzpicture}
        \node[gauge, label=below:$1$] (1l) at (0,0){};
        \node[gauge, label=below:$2$] (2l) at (-1,0){};
        \node[] (cdotsl) at (-2,0){$\cdots$};
        \node[gauge, label=below:$N$] (nl) at (-3,0){};
        \node[gauge, label=below:$N$] (nml) at (-4,0){};
        \node[] (cdotsm) at (-5,0){$\cdots$};
        \node[gauge, label=below:$N$] (nmr) at (-6,0){};
        \node[gauge, label=below:$N$] (nr) at (-7,0){};
        \node[gauge, label=below:$N-1$] (nm1) at (-8,0){};
        \node[] (cdotsr) at (-9,0){$\cdots$};
        \node[gauge, label=below:$2n+1$] (2rp3) at (-10,0){};
        \node[gauge, label=below:$2n$] (2rp2) at (-11,0){};
        
        \node[flavour, label=above:$1$] (1fl) at (-3,1){};
        \node[flavour, label=above:$1$] (1fr) at (-7,1){};

        \draw[-] (1l)--(2l)--(cdotsl)--(nl)--(nml)--(cdotsm)--(nmr)--(nr)--(nm1)--(cdotsr)--(2rp3)  (1fl)--(nl) (1fr)--(nr);

        \draw[transform canvas={yshift=1.3pt}] (2rp3)--(2rp2);
        \draw[transform canvas={yshift=-1.3pt}] (2rp3)--(2rp2);

        \draw[-] (-10.6,0.2)--(-10.4,0)--(-10.6,-0.2);

        \draw [decorate, decoration = {brace, raise=15pt, 
         amplitude=5pt}] (-3,0) --  (-7,0) node[pos=0.5,below=20pt,black]{$k+1$};
    \end{tikzpicture}
    \caption{}
    \label{eq:UnSOtheoryMirror}
    \end{subfigure}
    \begin{subfigure}[t]{0.75\textwidth}
    \centering
    \begin{tikzpicture}
        \draw[-] (0,-1)--(0,1) (-1,-1)--(-1,1) (-2,-1)--(-2,1) (-3,-1)--(-3,1) (-4,-1)--(-4,1) (-5,-1)--(-5,1) (-6,-1)--(-6,1) (-7,-1)--(-7,1) (-8,-1)--(-8,1) (-9,-1)--(-9,1) (-10,-1)--(-10,1) (-11,-1)--(-11,1) (-12,-1)--(-12,1);
        \draw[dashed] (-13,-1)--(-13,1)node[pos=1,above]{$\mathrm{ON}^+$};
        \draw[-] (0,0)--(-1,0)node[midway, above]{$1$}--(-2,0)node[midway, above]{$2$};
        \draw[-] (-2,0)--(-2.2,0) (-2.8,0)--(-3,0);
        \node at (-2.5,0){$\cdots$};
        \draw[-] (-3,0)--(-4,0)node[midway, above]{$N$};
        \node[D5] at (-3.5,0.75){};
        \draw[-] (-4,0)--(-5,0)node[midway, above]{$N$};
        \draw[-] (-5,0)--(-5.2,0) (-5.8,0)--(-6,0);
        \node at (-5.5,0){$\cdots$};
        \draw[-] (-6,0)--(-7,0)node[midway,above]{$N$}--(-8,0)node[midway,above]{$N$}--(-9,0)node[midway, above]{$N-1$};
        \node[D5] at (-7.5,0.75){};
        \draw[-] (-9,0)--(-9.2,0) (-9.8,0)--(-10,0);
        \node at (-9.5,0){$\cdots$};
        \draw[-] (-10,0)--(-11,0)node[midway, above]{$2n+2$}--(-12,0)node[midway,above]{$2n+1$}--(-13,0)node[midway, above]{$2n$};
        \draw [decorate, decoration = {brace, raise=5pt, 
         amplitude=5pt}] (-8,1) --  (-3,1) node[pos=0.5,above=10pt,black]{$k+2$};
    \end{tikzpicture}
    \caption{}
    \label{fig:UnSOMirrorBrane}
\end{subfigure}
    \caption{Gauging an $\sorm(2n)$ flavour subgroup of $\urm(N)$ SQCD with $2N+k$ flavours returns the theory in \Figref{eq:UnSOtheory}, realised from the brane system in \Figref{fig:UnSOBrane}. The magnetic theory is conjectured to be \Figref{eq:UnSOtheoryMirror}, associated to the brane system in \Figref{fig:UnSOMirrorBrane}, related to that of \Figref{fig:UnSOBrane} by brane-creation and S-duality. Comparison to the mirror of the SQCD in \eqref{quiv:UN_SQCD} leads to the conjectural quotient quiver \eqref{quiv:SO2n_quotient_quiver}.}
    \label{}
\end{figure}
The starting point for the analysis is once again the $\urm(N)$ SQCD with $2N+k$ flavours given in \eqref{quiv:UN_SQCD}. Gauging an $\sorm(2n)$ flavour symmetry subgroup (where $N\geq 2n+1$ ensures positive balance) results in the theory \Quiver{eq:UnSOtheory} given in \Figref{eq:UnSOtheory} with Type IIB brane system in \Figref{fig:UnSOBrane}. This theory has an $\surm(2N+k-2n)$ flavour symmetry -- the commutant of $\sorm(2n)$ inside $\surm(2N+k)$ -- and a Higgs branch of dimension $N(N+k)-n(2n+1)$. The magnetic theory is \Quiver{eq:UnSOtheoryMirror} given in \Figref{eq:UnSOtheoryMirror}, alongside its brane system in \Figref{fig:UnSOMirrorBrane}. Comparing the magnetic theory of the SQCD before (\eqref{quiv:UN_SQCD}) and after (\Figref{eq:UnSOtheoryMirror}) gauging identifies the quotient quiver \eqref{quiv:SO2n_quotient_quiver} and the subtraction prescription, given below.

Note that the total rank of the gauge nodes of the quotient quiver \eqref{quiv:SO2n_quotient_quiver} is $n(2n-1)$, the same as the dimension of $\sorm(2n)$.
\begin{equation}
\raisebox{-0.5\height}{\begin{tikzpicture}
    \node[] (A) at (-3,-0.1){$\sorm(2n) \; \text{Quotient Quiver}$};
    \node[gauge, label=below:$1$] (1l) at (0,0){};
    \node[gauge, label=below:$2$] (2l) at (1,0){};
    \node[] (cdots) at (2,0){$\cdots$};
    \node[gauge, label=above:$2n-2$] (2rm1) at (3,0){};
    \node[gauge, label=below:$2n-1$] (2r) at (4,0){};
    \draw[-] (1l)--(2l)--(cdots)--(2rm1)--(2r);
\end{tikzpicture}}
\label{quiv:SO2n_quotient_quiver}
\end{equation}
\paragraph{Formal statement of the rule}
The conjectured operation to gauge an $\sorm(2n)$ subgroup of the Coulomb branch global symmetry requires a long leg up to at least $\urm(2n)$ and involves the following steps. 
\begin{enumerate}
\item Take a target quiver with a long leg of gauge nodes up to at least $\urm(2n)$. As in Section \ref{sec:Sp(n)QQS}, the edge in \textcolor{teal}{teal} denotes some generic connection to the rest of the quiver $Q$.
\begin{equation}
\raisebox{-0.5\height}{\begin{tikzpicture}
    \node[gauge, label=below:$1$] (1) at (0,0){};
    \node[gauge, label=below:$2$] (2) at (1,0){};
    \node[] (cdots) (cdots) at (2,0){$\cdots$};
    \node[gauge, label=below:$2n-1$] (2r) at (3,0){};
    \node[gauge, label=below:$2n$] (2rp1) at (4,0){};
    \node[gauge] (Q) at (5,0){$Q$};
    \draw[-] (1)--(2)--(cdots)--(2r)--(2rp1);
    \draw[teal,very thick] (2rp1)--(Q);
    \end{tikzpicture}}  
\end{equation}
\item Align the $\sorm(2n)$ quotient quiver \eqref{quiv:SO2n_quotient_quiver} against the leg of the target quiver and subtract the ranks of the quotient quiver from those of the target. This deletes the gauge nodes from $\urm(1)$ up to $\urm(2n-1)$ in the maximal chain, leaving only $\urm(2n)$. Rebalancing is not required.
\begin{equation}
\raisebox{-0.5\height}{\begin{tikzpicture}
    \node[gauge, label=below:$1$] (1) at (0,0){};
    \node[gauge, label=below:$2$] (2) at (1,0){};
    \node[] (cdots) (cdots) at (2,0){$\cdots$};
    \node[gauge, label=below:$2n-1$] (2r) at (3,0){};
    \node[gauge, label=below:$2n$] (2rp1) at (4,0){};
    \node[gauge] (Q) at (5,0){$Q$};
    \draw[-] (1)--(2)--(cdots)--(2r)--(2rp1);
    \draw[teal,very thick] (2rp1)--(Q);
    \node[gauge, label=below:$1$] (1sub) at (0,-1){};
    \node[gauge, label=below:$2$] (2sub) at (1,-1){};
    \node[] (cdots) (cdotssub) at (2,-1){$\cdots$};
    \node[gauge, label=below:$2n-1$] (2rsub) at (3,-1){};
    \node[] (minus) at (-1,-1) {$-$};
    \draw[-] (1sub)--(2sub)--(cdotssub)--(2rsub);
    \node[gauge, label=below:$2n$] (2rp1r) at (4,-2){};
    \node[gauge] (Qr) at (5,-2){$Q$};
    \draw[teal,very thick] (2rp1r)--(Qr);    
    \end{tikzpicture}}
    \end{equation}
    \item All edges connected to the $\urm(2n)$ gauge node double their non-simply laced-ness -- the $\urm(2n)$ is the long root.
\begin{equation}
    \raisebox{-0.5\height}{\begin{tikzpicture}
    \node (a) at (0,0){\begin{tikzpicture}
    \node[gauge, label=below:$2n$] (2rp1) at (0,0){};
    \node[gauge] (Qr) at (1,0){$Q$};
    \draw[teal,very thick] (2rp1)--(Qr); 
    \end{tikzpicture}};
    \node (b) at (5,0){\begin{tikzpicture}
    \node[gauge, label=below:$2n$] (r) at (0,0){};
    \node[gauge] (Q) at (1,0){$Q$};
    \draw[teal, very thick, transform canvas={yshift=1.3pt}] (r)--(Q);
    \draw[teal, very thick, transform canvas={yshift=-1.3pt}] (r)--(Q);
    \draw[teal, very thick, -] (0.4,0.2)--(0.6,0)--(0.4,-0.2);
    \end{tikzpicture}};
    \draw[->] (a)--(b) node[midway, above]{\textrm{Lacing}};
    \end{tikzpicture}}
    \end{equation}
\end{enumerate}
Note that $\sorm(2n)$ quotient quiver subtraction requires a leg of gauge nodes $(1)-(2)-\cdots-(2n)-$, sourcing (at least) an $\surm(2n)$ factor in the Coulomb branch global symmetry. The $\sorm(2n)$ being gauged is embedded in this $\surm(2n)$ with discrete commutant, shown below. \begin{equation}
    [1,0,\cdots,0,1]_{\surm(2n)}\rightarrow [0,1,0,\cdots,0]_{\sorm(2n)}+[2,0,\cdots,0]_{\sorm(2n)}
\end{equation}
In the case that the target quiver has enhanced global symmetry, the $\sorm(2n)$ being gauged may be embedded directly into that global symmetry. Nevertheless, it remains sourced by the long leg.
\subsection{$\overline{min. F_4}///\sorm(2)$}
\begin{figure}[h!]
    \centering
    \begin{subfigure}{0.4\textwidth}
    \centering
    \begin{tikzpicture}
        \node[gauge, label=below:$1$] (1) at (0,0){};
        \node[gauge, label=below:$2$] (2) at (1,0){};
        \node[gauge, label=below:$3$] (3) at (2,0){};
        \node[gauge, label=below:$2$] (2r) at (3,0){};
        \node[gauge, label=below:$1$] (1r) at (4,0){};

        \draw[-] (1)--(2)--(3) (2r)--(1r);
        \draw[transform canvas={yshift=1.3pt}](3)--(2r);
        \draw[transform canvas={yshift=-1.3pt}](3)--(2r);
        \draw[-] (2.4,0.2)--(2.6,0)--(2.4,-0.2);
    
         \node[gauge, label=below:$1$] (1sub) at (0,-1){};
        \node[] (minus) at (-1,-1) {$-$};

        \node[gauge, label=below:$2$] (2res) at (1,-2){};
        \node[gauge, label=below:$3$] (3res) at (2,-2){};
        \node[gauge, label=below:$2$] (2rres) at (3,-2){};
        \node[gauge, label=below:$1$] (1rres) at (4,-2){};

        \draw[-] (2res)--(3res) (2rres)--(1rres);
        \draw[transform canvas={yshift=1.3pt}](3res)--(2rres);
        \draw[transform canvas={yshift=-1.3pt}](3res)--(2rres);
        \draw[-] (2.4,-1.8)--(2.6,-2)--(2.4,-2.2);
    \end{tikzpicture}     
    \caption{}
    \label{fig:F4SO2Sub}
    \end{subfigure}
    \begin{subfigure}{0.4\textwidth}
    \centering
    \begin{tikzpicture}
    \node (a) at (0,0){\begin{tikzpicture}
        \node[gauge, label=below:$2$] (2res) at (1,-2){};
        \node[gauge, label=below:$3$] (3res) at (2,-2){};
        \node[gauge, label=below:$2$] (2rres) at (3,-2){};
        \node[gauge, label=below:$1$] (1rres) at (4,-2){};
        \draw[-] (2res)--(3res) (2rres)--(1rres);
        \draw[transform canvas={yshift=1.3pt}](3res)--(2rres);
        \draw[transform canvas={yshift=-1.3pt}](3res)--(2rres);
        \draw[-] (2.4,-1.8)--(2.6,-2)--(2.4,-2.2);
    \end{tikzpicture}};
        \node (b) at (0,-3){\begin{tikzpicture}
        \node[gauge, label=below:$2$] (2res) at (1,-2){};
        \node[gauge, label=below:$3$] (3res) at (2,-2){};
        \node[gauge, label=below:$2$] (2rres) at (3,-2){};
        \node[gauge, label=below:$1$] (1rres) at (4,-2){};

        \draw[-] (3res) (2rres)--(1rres);
        \draw[transform canvas={yshift=1.3pt}](3res)--(2rres);
        \draw[transform canvas={yshift=-1.3pt}](3res)--(2rres);
        \draw[-] (2.4,-1.8)--(2.6,-2)--(2.4,-2.2);
        \draw[transform canvas={yshift=1.3pt}](2res)--(3res);
        \draw[transform canvas={yshift=-1.3pt}](2res)--(3res);
        \draw[-] (1.4,-1.8)--(1.6,-2)--(1.4,-2.2);
        \end{tikzpicture}};
        \draw[->] (a)--(b) node[midway, right]{\textrm{Lacing}};
        \end{tikzpicture}
    \caption{}
    \label{fig:F4SO2Lace}
    \end{subfigure}
    \begin{subfigure}{0.16\textwidth}
    \centering
    \begin{tikzpicture}
    \node[hasse] (1) at (0,0){};
    \node[hasse] (2) at (0,-1.5){};
    \node[hasse] (3) at (0,-3){};
    \draw[-] (1)--(2)node[midway, right]{$a_4$}--(3) node[midway, right]{$c_3$};
    \end{tikzpicture}
    \caption{}
    \label{fig:F4SO2CoulHasse}
    \end{subfigure}
    \caption{$\sorm(2)$ \hyperref[fig:SO2nQQS]{quotient quiver subtraction} on the affine $F^{(1)}_4$ quiver occurs in two steps. The first is shown in \ref{fig:F4SO2Sub} and involves the subtraction of the $\sorm(2)$ quotient quiver from the maximal leg. The second step, given in \ref{fig:F4SO2Lace}, is the non-simple lacing of the edge connecting the left $\urm(2)$ node to the $\urm(3)$, producing the resulting quiver \Quiver{fig:F4SO2QQS}. The Coulomb branch Hasse diagram is given in \Figref{fig:F4SO2CoulHasse}.}
    \label{fig:F4SO2QQS}
\end{figure}
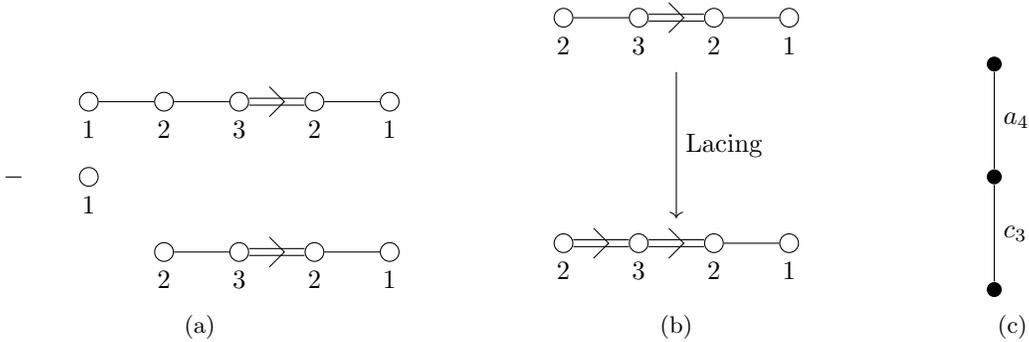

The affine $F_4^{(1)}$ quiver has Coulomb branch $\overline{min. F_4}$. It contains a tail of gauge nodes on the long side of the quiver which goes as $(1)-(2)-(3)-$ and is suitable for $\sorm(2)$ \hyperref[fig:SO2nQQS]{quotient quiver subtraction}. This operation is shown in \Figref{fig:F4SO2QQS} and produces \Quiver{fig:F4SO2QQS}, with the subtraction of the quotient quiver in \Figref{fig:F4SO2Sub} and the non-simple lacing in \Figref{fig:F4SO2Lace}.

The refined Coulomb branch Hilbert series of \Figref{fig:F4SO2QQS} is cumbersome. However, the unrefined Coulomb branch Hilbert series is \begin{equation}
    \hsC{fig:F4SO2QQS}=\frac{\left(\begin{aligned}1 &+ 14 t^2 + 168 t^4 + 854 t^6 + 3136 t^8 + 7062 t^{10} + 11991 t^{12} + 
 13708 t^{14} \\&+ 11991 t^{16} + 7062 t^{18} + 3136 t^{20} + 854 t^{22} + 
 168 t^{24} + 14 t^{26} + t^{28}\end{aligned}\right)}{(1 - t^2)^7 (1 - t^4)^7}
\end{equation}Although this moduli space has no particular name, its $\sprm(3)$ symmetry can be confirmed from the refined Hilbert series (not presented here). The Coulomb branch Hasse diagram is also straightforwardly computed and given in \Figref{fig:F4SO2CoulHasse}. It is still unclear how to compute the Higgs branch Hasse diagram for non-simply laced quivers \cite{Bennett:2024loi}.

The conclusion is that \begin{equation}
    \overline{min. F_4}///\sorm(2)=\mathcal C\left(\text{\Quiver{fig:F4SO2QQS}}\right)
\end{equation}which is further verified from an explicit Weyl integration.

The following embedding of $F_4\hookleftarrow \sprm(3)\times\urm(1)$ is employed where the fundamental representation of $F_4$ decomposes as \begin{equation}
    \left(\mu_1\right)_{F_4}\rightarrow \mu_1\left(q+1/q\right)+\mu_2
\end{equation}where the $\mu_{1,2}$ on the right hand side refer to $\sprm(3)$ highest weight fugacities and $q$ is a $\urm(1)$ fugacity.

\subsection{$\overline{min. E_7}///\sorm(4)$}
\label{sec:E7_so4}
\begin{figure}[h!]
    \centering
    \begin{subfigure}{0.49\textwidth}
    \begin{tikzpicture}
        \node[gauge, label=below:$1$] (1l) at (0,0){};
        \node[gauge, label=below:$2$] (2l) at (1,0){};
        \node[gauge, label=below:$3$] (3l) at (2,0){};
        \node[gauge, label=below:$4$] (4) at (3,0){};
        \node[gauge, label=below:$3$] (3r) at (4,0){};
        \node[gauge, label=below:$2$] (2r) at (5,0){};
        \node[gauge, label=below:$1$] (1r) at (6,0){};
        \node[gauge, label=above:$2$] (2t) at (3,1){};
        \draw[-] (1l)--(2l)--(3l)--(4)--(3r)--(2r)--(1r) (2t)--(4);
        \node[gauge, label=below:$1$] (1sub) at (0,-1){};
        \node[gauge, label=below:$2$] (2sub) at (1,-1){};
        \node[gauge, label=below:$3$] (3sub) at (2,-1){};
        \node[] (minus) at (-1,-1) {$-$};
        \draw[-] (1sub)--(2sub)--(3sub);
        \node[gauge, label=above:$2$] (2tres) at (3,-2){};
        \node[gauge, label=below:$4$] (4res) at (3,-3){};
        \node[gauge, label=below:$3$] (3res) at (4,-3){};
        \node[gauge, label=below:$2$]  (2res) at (5,-3){};
        \node[gauge, label=below:$1$] (1res) at (6,-3){};
        \draw[-] (2tres)--(4res)--(3res)--(2res)--(1res);
    \end{tikzpicture}     
    \caption{}
    \label{fig:E7SO4Sub}
    \end{subfigure}
    \begin{subfigure}{0.49\textwidth}
    \centering
        \begin{tikzpicture}
            \node (a) at (0,0){\begin{tikzpicture}
               \node[gauge, label=below:$2$] (2tres) at (2,-3){};
        \node[gauge, label=below:$4$] (4res) at (3,-3){};
        \node[gauge, label=below:$3$] (3res) at (4,-3){};
        \node[gauge, label=below:$2$]  (2res) at (5,-3){};
        \node[gauge, label=below:$1$] (1res) at (6,-3){};
        \draw[-] (2tres)--(4res)--(3res)--(2res)--(1res);
            \end{tikzpicture}};
        \node (b) at (0,-3){\begin{tikzpicture}
        \node[gauge, label=below:$2$] (2tres) at (2,-3){};
        \node[gauge, label=below:$4$] (4res) at (3,-3){};
        \node[gauge, label=below:$3$] (3res) at (4,-3){};
        \node[gauge, label=below:$2$]  (2res) at (5,-3){};
        \node[gauge, label=below:$1$] (1res) at (6,-3){};
        \draw[-] (3res)--(2res)--(1res);

        \draw[transform canvas={yshift=1.3pt}](2tres)--(4res);
        \draw[transform canvas={yshift=-1.3pt}](2tres)--(4res);
        \draw[-] (2.6,-3.2)--(2.4,-3)--(2.6,-2.8);
        \draw[transform canvas={yshift=1.3pt}](4res)--(3res);
        \draw[transform canvas={yshift=-1.3pt}](4res)--(3res);
        \draw[-] (3.4,-3.2)--(3.6,-3)--(3.4,-2.8);
        \end{tikzpicture}};
        \draw[->] (a)--(b) node[midway, right]{\textrm{Lacing}};
        \end{tikzpicture}
\caption{}
\label{fig:E7SO4Lace}
\end{subfigure}
\caption{$\sorm(4)$ \hyperref[fig:SO2nQQS]{quotient quiver subtraction} on the affine $E^{(1)}_7$ quiver occurs in two steps. The first step, shown in \ref{fig:E7SO4Sub}, involves the subtraction of the $\sorm(4)$ quotient quiver from one maximal leg. The second step, in \ref{fig:E7SO4Lace}, is the non-simple lacing of the two edges emerging from the $\urm(4)$ node.}
\label{fig:E7SO4QQS}
\end{figure}
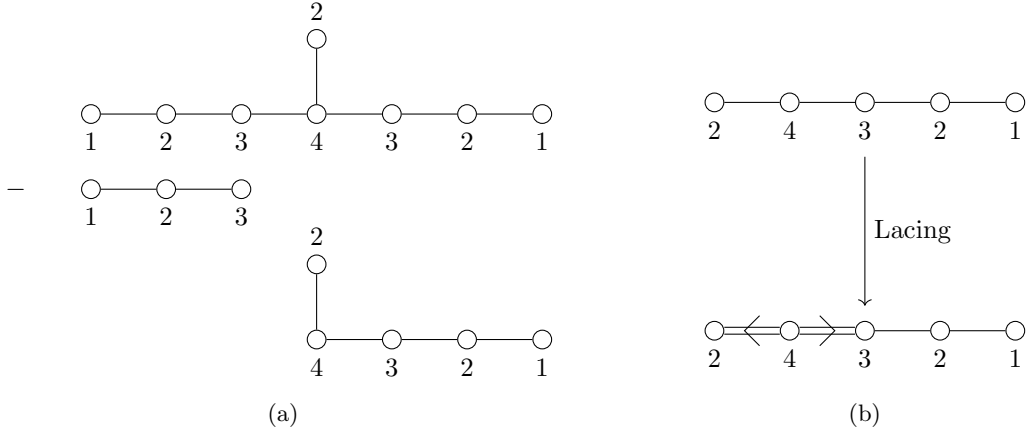

The affine $E_7^{(1)}$ quiver has Coulomb branch $\overline{min. E_7}$. Its two maximal legs of $(1)-(2)-(3)-(4)-$ each support an $\sorm(4)$ \hyperref[fig:SO2nQQS]{quotient quiver subtraction}. One such subtraction, shown explicitly in \Figref{fig:E7SO4QQS}, produces \Quiver{fig:E7SO4QQS}. Although its refined Coulomb branch Hilbert series is cumbersome, the unrefined expression is
\begin{equation}
    \hsC{fig:E7SO4QQS}=\frac{\left(\begin{aligned}1 &+ 7 t^2 + 100 t^4 + 596 t^6 + 
  3531 t^8 + 13712 t^{10} + 45536 t^{12}+ 112650 t^{14} \\&+ 233881 t^{16} + 
  380039 t^{18} + 519199 t^{20} + 565768 t^{22} + \cdots + t^{44}\end{aligned}\right)}{(1 - t^2)^{11} (1 - t^4)^{11}}
\end{equation}
This moduli space, which has no particular name, has $\sorm(6)\times\surm(2)$ symmetry, confirmed by the refined Hilbert series.

The same moduli space arises from Weyl integration and the embedding $E_7\hookleftarrow D_6\times A_1\hookleftarrow D_3\times D_3\times A_1\hookleftarrow D_2\times D_3\times A_1$ which decomposes the $[0,0,0,0,0,1,0]$ of $E_7$ as \begin{equation}
    \left(\mu_6\right)_{E_7}\rightarrow \left(\mu_1+\nu_1^2+\nu_2^2\right)\rho+\left(\mu_2+\mu_3\right)\nu_1\nu_2
\end{equation} 
Note that the $\mu, \nu,$ and $\rho$ are highest weight fugacities for $D_3,\;D_2,\;$and $A_1$ respectively. The integration is performed with respect to the $D_2$ fugacities.
\subsection{$\overline{min. E_8}///\sorm(4)$}
\label{sec:E8_so4}
Consider again the affine $E_8^{(1)}$ quiver, whose maximal leg $(1)-(2)-\cdots-(6)-$ admits an $\sorm(4)$ \hyperref[fig:SO2nQQS]{quotient quiver subtraction}. This operation, shown in \Figref{fig:E8SO4QQS}, produces \Quiver{fig:E8SO4QQS}.

The refined Coulomb branch Hilbert series of \Figref{fig:E8SO4QQS} is cumbersome -- its unrefined form is \begin{equation}
    \hsC{fig:E8SO4QQS}=\frac{\left(\begin{aligned}1 &+ 22 t^2 + 582 t^4 + 10360 t^6 + 
   154981 t^8 + 1822118 t^{10} + 17548922 t^{12} + 139175916 t^{14} \\&+ 
   925774625 t^{16} + 5219073951 t^{18} + 25226111030 t^{20} + 
   105435301585 t^{22} \\&+ 384139161711 t^{24} + 1227840826079 t^{26} + 
   3462893979058 t^{28} + 8658011097494 t^{30} \\&+ 19269044472299 t^{32} + 
   38301892953919 t^{34} + 68192783223365 t^{36} \\&+ 108993744295329 t^{38} + 
   156680149569483 t^{40} + 202847214133311 t^{42} \\&+ 
   236754481344743 t^{44} + 249255260007352 t^{46} +\cdots + t^{92}\end{aligned}\right)}{(1 - t^2)^{23} (1 - t^4)^{23}} 
\end{equation}
This moduli space, which has no particular name as a symplectic singularity, has $\sorm(10)$ global symmetry.
\begin{figure}[h!]
\centering
\begin{subfigure}{0.49\textwidth}
\begin{tikzpicture}
        \node[gauge, label=below:$1$] (1l) at (0,0){};
        \node[gauge, label=below:$2$] (2l) at (1,0){};
        \node[gauge, label=below:$3$] (3l) at (2,0){};
        \node[gauge, label=below:$4$] (4l) at (3,0){};
        \node[gauge, label=below:$5$] (5l) at (4,0){};
        \node[gauge, label=below:$6$] (6) at (5,0){};
        \node[gauge, label=below:$4$] (4r) at (6,0){};
        \node[gauge, label=below:$2$] (2r) at (7,0){};
        \node[gauge, label=above:$3$] (3t) at (5,1){};
        \draw[-] (1l)--(2l)--(3l)--(4l)--(5l)--(6)--(4r)--(2r) (6)--(3t);
        \node[gauge, label=below:$1$] (1sub) at (0,-1){};
        \node[gauge, label=below:$2$] (2sub) at (1,-1){};
        \node[gauge, label=below:$3$] (3sub) at (2,-1){};
        \node[] (minus) at (-1,-1) {$-$};
        \draw[-] (1sub)--(2sub)--(3sub);
        \node[gauge, label=above:$3$] (3tres) at (5,-2){};
        \node[gauge, label=below:$6$] (6res) at (5,-3){};
        \node[gauge, label=below:$4$] (4rres) at (6,-3){};
        \node[gauge, label=below:$2$]  (2rres) at (7,-3){};
        \node[gauge, label=below:$5$] (5res) at (4,-3){};
        \node[gauge, label=below:$4$] (4res) at (3,-3){};
        \draw[-] (4res)--(5res)--(6res)--(4rres)--(2rres) (3tres)--(6res);
    \end{tikzpicture}     
    \caption{}
    \label{fig:E8SO4Sub}
    \end{subfigure}
    \begin{subfigure}{0.49\textwidth}
    \centering
    \begin{tikzpicture}
    \node (a) at (0,0){\begin{tikzpicture}
        \node[gauge, label=above:$3$] (3tres) at (5,-2){};
        \node[gauge, label=below:$6$] (6res) at (5,-3){};
        \node[gauge, label=below:$4$] (4rres) at (6,-3){};
        \node[gauge, label=below:$2$]  (2rres) at (7,-3){};
        \node[gauge, label=below:$5$] (5res) at (4,-3){};
        \node[gauge, label=below:$4$] (4res) at (3,-3){};
        \draw[-] (4res)--(5res)--(6res)--(4rres)--(2rres) (3tres)--(6res);
        \end{tikzpicture}};
    \node (b) at (0,-3){\begin{tikzpicture}
        \node[gauge, label=above:$3$] (3tres) at (5,-2){};
        \node[gauge, label=below:$6$] (6res) at (5,-3){};
        \node[gauge, label=below:$4$] (4rres) at (6,-3){};
        \node[gauge, label=below:$2$]  (2rres) at (7,-3){};
        \node[gauge, label=below:$5$] (5res) at (4,-3){};
        \node[gauge, label=below:$4$] (4res) at (3,-3){};
        \draw[-] (5res)--(6res)--(4rres)--(2rres) (3tres)--(6res);
        \draw[transform canvas={yshift=1.3pt}](4res)--(5res);
        \draw[transform canvas={yshift=-1.3pt}](4res)--(5res);
        \draw[-] (3.4,-3.2)--(3.6,-3)--(3.4,-2.8);
        \end{tikzpicture}};
        \draw[->] (a)--(b) node[midway, right]{\textrm{Lacing}};
    \end{tikzpicture}
    \caption{}
    \label{fig:E8SO4Lace}
    \end{subfigure}
    \caption{$\sorm(4)$ \hyperref[fig:SO2nQQS]{quotient quiver subtraction} on the affine $E^{(1)}_8$ quiver occurs in two steps. The first step is shown in \ref{fig:E8SO4Sub} and involves the subtraction of the $\sorm(4)$ quotient quiver from the maximal leg. The second step, shown \ref{fig:E8SO4Lace}, is the non-simple lacing of the two edges going from the $\urm(4)$ node producing  \Quiver{fig:E8SO4QQS}.}
    \label{fig:E8SO4QQS}
\end{figure}

Weyl integration straightforwardly constructs the same moduli space under the embedding $E_8\hookleftarrow D_8\hookleftarrow\times A_3\hookleftarrow D_5\hookleftarrow D_2\times D_5$, which decomposes the adjoint of $E_8$ as \begin{equation}
    \left(\mu_7\right)_{E_8}\rightarrow \mu_2 + (1 + \mu_1) \nu_1^2 + (\mu_4 + \mu_5) \nu_1 \nu_2 + (1 + \mu_1) \nu_2^2 + \nu_1^2 \nu_2^2
\end{equation}The $\mu$ and $\nu$ are highest weight fugacities for $D_5$ and $D_2$ respectively and the integration is performed with respect to the $D_2$ fugacities.
\section{\boldmath$\sorm(2n+1)$\;\unboldmath Quotient Quivers}
\label{sec:So(2n+1)QQS}
\begin{figure}
    \centering
    \begin{subfigure}[t]{0.4\textwidth}
        \begin{tikzpicture}
        \node[gauge, label=below:$N$] (n)at (0,0){};
        \node[gauger, label=below:$\sorm(2n+1)$] (spr) at (-1,0){};
        \node[flavour, label=above:$2N+k-2n-1$] (flav) at (0,1){};
        \draw[-] (spr)--(n)--(flav);
    \end{tikzpicture}
    \caption{}
    \label{eq:UnBtheory}
    \end{subfigure}
    \begin{subfigure}[t]{0.4\textwidth}
        \begin{tikzpicture}
        \draw[-] (0,-1)--(0,1) (-2,-1)--(-2,1);
        \node[O5circle,label=below:$\widetilde{\mathrm{O5}^+}$] at (-4,0){};
        \draw[-] (0,0)--(-2,0)node[midway, above]{$N$}--(-3.8,0)node[midway, above]{$2n+1$};
        \node[D5] at (-0.5,0.75){};
        \node[D5] at (-1.5,0.75){};
        \node at (-1,0.75){$\cdots$};
        \draw [decorate, decoration = {brace, raise=10pt, 
         amplitude=5pt}] (-1.5,0.75) --  (-0.5,0.75) node[pos=0.5,above=15pt,black]{$2N+k-2n-1$};  
    \end{tikzpicture}
    \caption{}
    \label{fig:UnBBrane}
    \end{subfigure}
    \begin{subfigure}[t]{0.75\textwidth}
    \begin{tikzpicture}
        \node[gauge, label=below:$1$] (1l) at (0,0){};
        \node[gauge, label=below:$2$] (2l) at (-1,0){};
        \node[] (cdotsl) at (-2,0){$\cdots$};
        \node[gauge, label=below:$N$] (nl) at (-3,0){};
        \node[gauge, label=below:$N$] (nml) at (-4,0){};
        \node[] (cdotsm) at (-5,0){$\cdots$};
        \node[gauge, label=below:$N$] (nmr) at (-6,0){};
        \node[gauge, label=below:$N$] (nr) at (-7,0){};
        \node[gauge, label=below:$N-1$] (nm1) at (-8,0){};
        \node[] (cdotsr) at (-9,0){$\cdots$};
        \node[gauge, label=below:$2n+2$] (2rp3) at (-10,0){};
        \node[gauge, label=below:$2n+1$] (2rp2) at (-11,0){};
        
        \node[flavour, label=above:$1$] (1fl) at (-3,1){};
        \node[flavour, label=above:$1$] (1fr) at (-7,1){};

        \draw[-] (1l)--(2l)--(cdotsl)--(nl)--(nml)--(cdotsm)--(nmr)--(nr)--(nm1)--(cdotsr)--(2rp3)  (1fl)--(nl) (1fr)--(nr);

        \draw[transform canvas={yshift=1.3pt}] (2rp3)--(2rp2);
        \draw[transform canvas={yshift=-1.3pt}] (2rp3)--(2rp2);

        \draw[-] (-10.6,0.2)--(-10.4,0)--(-10.6,-0.2);

        \draw [decorate, decoration = {brace, raise=15pt, 
         amplitude=5pt}] (-3,0) --  (-7,0) node[pos=0.5,below=20pt,black]{$k+1$};
    \end{tikzpicture}
    \caption{}
    \label{eq:UnBtheoryMirror}
    \end{subfigure}
    \begin{subfigure}[t]{0.75\textwidth}
    \centering
    \begin{tikzpicture}
        \draw[-] (0,-1)--(0,1) (-1,-1)--(-1,1) (-2,-1)--(-2,1) (-3,-1)--(-3,1) (-4,-1)--(-4,1) (-5,-1)--(-5,1) (-6,-1)--(-6,1) (-7,-1)--(-7,1) (-8,-1)--(-8,1) (-9,-1)--(-9,1) (-10,-1)--(-10,1) (-11,-1)--(-11,1) (-12,-1)--(-12,1);
        \draw[dashed] (-13,-1)--(-13,1)node[pos=1,above]{$\widetilde{\mathrm{ON}^+}$};
        \draw[-] (0,0)--(-1,0)node[midway, above]{$1$}--(-2,0)node[midway, above]{$2$};
        \draw[-] (-2,0)--(-2.2,0) (-2.8,0)--(-3,0);
        \node at (-2.5,0){$\cdots$};
        \draw[-] (-3,0)--(-4,0)node[midway, above]{$N$};
        \node[D5] at (-3.5,0.75){};
        \draw[-] (-4,0)--(-5,0)node[midway, above]{$N$};
        \draw[-] (-5,0)--(-5.2,0) (-5.8,0)--(-6,0);
        \node at (-5.5,0){$\cdots$};
        \draw[-] (-6,0)--(-7,0)node[midway,above]{$N$}--(-8,0)node[midway,above]{$N$}--(-9,0)node[midway, above]{$N-1$};
        \node[D5] at (-7.5,0.75){};
        \draw[-] (-9,0)--(-9.2,0) (-9.8,0)--(-10,0);
        \node at (-9.5,0){$\cdots$};
        \draw[-] (-10,0)--(-11,0)node[midway, above]{$2n+3$}--(-12,0)node[midway,above]{$2n+2$}--(-13,0)node[midway, above]{$2n+1$};
        \draw [decorate, decoration = {brace, raise=5pt, 
         amplitude=5pt}] (-8,1) --  (-3,1) node[pos=0.5,above=10pt,black]{$k+2$};
    \end{tikzpicture}
    \caption{}
    \label{fig:UnBMirrorBrane}
\end{subfigure}
    \caption{Gauging an $\sorm(2n+1)$ flavour subgroup of $\urm(N)$ SQCD with $2N+k$ flavours returns the theory in \Figref{eq:UnBtheory}, realised from the brane system in \Figref{fig:UnBBrane}. The magnetic theory is conjectured to be \Figref{eq:UnBtheoryMirror}, associated to the brane system in \Figref{fig:UnBMirrorBrane}, related to that of \Figref{fig:UnBBrane} by brane-creation and S-duality. Comparison to the mirror of the SQCD in \eqref{quiv:UN_SQCD} leads to the conjectural quotient quiver \eqref{quiv:Bn_quotient_quiver}.}
    \label{}
\end{figure}
The starting point for the analysis is once again the $\urm(N)$ SQCD with $2N+k$ flavours given in \eqref{quiv:UN_SQCD}. Gauging an $\sorm(2n+1)$ flavour symmetry subgroup results in the theory \Quiver{eq:UnBtheory} given in \Figref{eq:UnBtheory} with Type IIB brane system in \Figref{fig:UnBBrane}. The magnetic theory is \Quiver{eq:UnBtheoryMirror} given in \Figref{eq:UnBtheoryMirror}, alongside its brane system in \Figref{fig:UnBMirrorBrane}. Comparing the magnetic theory of the SQCD before (\eqref{quiv:UN_SQCD}) and after (\Figref{eq:UnBtheoryMirror}) gauging identifies the quotient quiver \eqref{quiv:Bn_quotient_quiver} and the subtraction prescription, given below. The total rank of gauge nodes in the quotient quiver is $n(2n+1)=\textrm{dim}\left(\sorm(2n+1)\right)$.
\begin{equation}
    \raisebox{-0.5\height}{\begin{tikzpicture}
    \node[] (A) at (-3,-0.1){$\sorm(2n+1) \; \text{Quotient Quiver}$};
        \node[gauge, label=below:$1$] (1) at (0,0){};
        \node[gauge, label=below:$2$] (2) at (1,0){};
        \node[] (cdots) at (2,0){$\cdots$};
        \node[gauge, label=below:$2n-1$] (2nm1) at (3,0){};
        \node[gauge, label=below:$2n$] (2n) at (4,0){};
        \draw[-] (1)--(2)--(cdots)--(2nm1)--(2n);
    \end{tikzpicture}}
\label{quiv:Bn_quotient_quiver}
\end{equation}
The gauging algorithm, which is almost identical to the $\sorm(2n)$ case, is given below for completeness.
\paragraph{Formal Statement of the Rule}
\begin{enumerate}
    \item Take a target quiver with a long leg of gauge nodes up to at least $\urm(2n+1)$. The edge in \textcolor{teal}{teal} denotes some generic connection to the rest of the quiver $\mathcal Q$ -- the only restriction is that the long leg must correspond to long roots of the Coulomb branch global symmetry algebra.
    \begin{equation}
    \begin{tikzpicture}
    \node[gauge, label=below:$1$] (1) at (0,0){};
    \node[gauge, label=below:$2$] (2) at (1,0){};
    \node[] (cdots) (cdots) at (2,0){$\cdots$};
    \node[gauge, label=below:$2n$] (2r) at (3,0){};
    \node[gauge, label=below:$2n+1$] (2rp1) at (4,0){};
    \node[gauge] (Q) at (5,0){$Q$};
    \draw[-] (1)--(2)--(cdots)--(2r)--(2rp1);
    \draw[teal,very thick] (2rp1)--(Q);
    \node[gauge, label=below:$1$] (1sub) at (0,-1){};
    \node[gauge, label=below:$2$] (2sub) at (1,-1){};
    \node[] (cdots) (cdotssub) at (2,-1){$\cdots$};
    \node[gauge, label=below:$2n$] (2rsub) at (3,-1){};
    \draw[-] (1sub)--(2sub)--(cdotssub)--(2rsub);      
    \end{tikzpicture}  
    \end{equation}
    \item Align end of the $\sorm(2n+1)$ quotient quiver against the leg of the target quiver and subtract the ranks of the quotient quiver from those of the target. This deletes the gauge nodes from $\urm(1)$ up to $\urm(2n)$ in the maximal chain, leaving only $\urm(2n+1)$. Rebalancing is not required.
    \begin{equation}
    \begin{tikzpicture}
    \node[gauge, label=below:$1$] (1) at (0,0){};
    \node[gauge, label=below:$2$] (2) at (1,0){};
    \node[] (cdots) (cdots) at (2,0){$\cdots$};
    \node[gauge, label=below:$2n$] (2r) at (3,0){};
    \node[gauge, label=below:$2n+1$] (2rp1) at (4,0){};
    \node[gauge] (Q) at (5,0){$Q$};
    \draw[-] (1)--(2)--(cdots)--(2r)--(2rp1);
    \draw[teal,very thick] (2rp1)--(Q);
    \node[gauge, label=below:$1$] (1sub) at (0,-1){};
    \node[gauge, label=below:$2$] (2sub) at (1,-1){};
    \node[] (cdots) (cdotssub) at (2,-1){$\cdots$};
    \node[gauge, label=below:$2n$] (2rsub) at (3,-1){};
    \node[] (minus) at (-1,-1) {$-$};
    \draw[-] (1sub)--(2sub)--(cdotssub)--(2rsub);
    \node[gauge, label=below:$2n+1$] (2rp1r) at (4,-2){};
    \node[gauge] (Qr) at (5,-2){$Q$};
    \draw[teal,very thick] (2rp1r)--(Qr);      
    \end{tikzpicture}  
    \end{equation}
    \item All edges connected to the $\urm(2n+1)$ gauge node (shown in \textcolor{teal}{teal}) double their non-simply laced-ness, with the $\urm(2n+1)$ corresponding to the long root.
    \begin{equation}
    \begin{tikzpicture}
    \node (a) at (0,0){\begin{tikzpicture}
    \node[gauge, label=below:$2n+1$] (2rp1) at (0,0){};
    \node[gauge] (Qr) at (1,0){$Q$};
    \draw[teal,very thick] (2rp1)--(Qr); 
    \end{tikzpicture}};
    \node (b) at (5,0){\begin{tikzpicture}
    \node[gauge, label=below:$2n+1$] (r) at (0,0){};
    \node[gauge] (Q) at (1,0){$Q$};
    \draw[teal, very thick, transform canvas={yshift=1.3pt}] (r)--(Q);
    \draw[teal, very thick, transform canvas={yshift=-1.3pt}] (r)--(Q);
    \draw[teal, very thick, -] (0.4,0.2)--(0.6,0)--(0.4,-0.2);
    \end{tikzpicture}};
    \draw[->] (a)--(b) node[midway, above]{\textrm{Lacing}};
    \end{tikzpicture}
    \end{equation}
\end{enumerate}

Note that $\sorm(2n+1)$ quotient quiver subtraction requires a $(1)-(2)-\cdots-(2n+1)-$ long leg of gauge nodes, which sources (at least) an $\surm(2n+1)$ factor of the Coulomb branch global symmetry. The $\sorm(2n+1)$ to be gauged is embedded with discrete commutant as follows.
\begin{equation}
    [1,0,\cdots,0,1]_{\surm(2n+1)}\rightarrow [0,1,0,\cdots,0]_{\sorm(2n+1)}+[2,0,\cdots,0]_{\sorm(2n+1)}
\end{equation}
This is exactly analogous to the embedding used for $\sorm(2n)$ quotient quiver subtraction. If the target quiver has enhanced global symmetry, the $\sorm(2n+1)$ being gauged may be embedded directly into that global symmetry. The origin remains in the gauge nodes in the long leg. 

It is important to address the various degeneracies inspired by accidental isomorphisms. The $A_1\cong B_1\cong C_1$ isomorphism is particularly interesting as it establishes that $\surm(2)$ and $\sprm(1)$ quotient quiver subtraction are identical operations. Interestingly, $\sorm(3)$ quotient quiver subtraction differs, and the $B_1$ embedding in the global symmetry of the long leg is distinct to that for $A_1$ and $C_1$. Explicitly, $\sorm(3)$ is embedded inside an $\surm(3)$ of the long leg whereas $\surm(2)$ and $\sprm(1)$ are embedded inside $\surm(2)$. For similar reasons the $B_2\cong C_2$ isomorphism does not extend to $\sorm(5)$ and $\sprm(2)$ quotient quiver subtraction, both because the resulting quivers are different and because $\sorm(5)$ is embedded inside $\surm(5)$ and $\sprm(2)$ is embedded inside $\surm(4)$.
\subsection{$\overline{min. E_6}///\sorm(3)$}
The affine $E_6^{(1)}$ quiver's $(1)-(2)-(3)-$ long leg is amenable to $\sorm(3)$ \hyperref[fig:SO2np1QQS]{quotient quiver subtraction}. This is shown in \Figref{fig:E6SO3QQS} and returns \Quiver{fig:E6SO3QQS}. The refined Coulomb branch Hilbert series is unwieldy -- the unrefined Coulomb branch Hilbert series is \begin{equation}
    \hsC{fig:E6SO3QQS}=\frac{\left(\begin{aligned}1 &+ 8 t^2 + 99 t^4 + 479 t^6 + 2095 t^8 + 5355 t^{10} + 11577 t^{12} + 
 16622 t^{14} + 20072 t^{16} \\&+ 16622 t^{18} + 11577 t^{20} + 5355 t^{22} + 
 2095 t^{24} + 479 t^{26} + 99 t^{28} + 
 8 t^{30} + t^{32}\end{aligned}\right)}{(1 - t^2)^{8} (1 - t^4)^8}
\end{equation}This moduli space, which has no particular name as a symplectic singularity, has $\surm(3)\times\surm(3)$ symmetry,  confirmed at the level of the refined Hilbert series.
\begin{figure}[h!]
    \centering
    \begin{subfigure}{0.45\linewidth}
    \centering
    \begin{tikzpicture}
        \node[gauge, label=below:$1$] (1l) at (0,0){};
        \node[gauge, label=below:$2$] (2l) at (1,0){};
        \node[gauge, label=below:$3$] (3) at (2,0){};
        \node[gauge, label=below:$2$] (2r) at (3,0){};
        \node[gauge, label=below:$1$] (1r) at (4,0){};
        \node[gauge, label=right:$2$] (2t) at (2,1){};
        \node[gauge, label=above:$1$] (1t) at (2,2){};
        \draw[-] (1l)--(2l)--(3)--(2r)--(1r) (3)--(2t)--(1t);
        \node[gauge, label=below:$1$] (1sub) at (0,-1){};
        \node[gauge, label=below:$2$] (2sub) at (1,-1){};
        \node[] (minus) at (-1,-1) {$-$};
        \draw[-] (1sub)--(2sub);
        \node[gauge, label=above:$1$] (1tres) at (2,-2){};
        \node[gauge, label=right:$2$] (2tres) at (2,-3){};
        \node[gauge, label=below:$3$] (3res) at (2,-4){};
        \node[gauge, label=below:$2$] (2rres) at (3,-4){};
        \node[gauge, label=below:$1$] (1rres) at (4,-4){};
        \draw[-] (1tres)--(2tres)--(3res)--(2rres)--(1rres);
    \end{tikzpicture}     
    \caption{}
    \label{fig:E6SO3Sub}
    \end{subfigure}
    \begin{subfigure}{0.45\linewidth}
    \centering
    \begin{tikzpicture}
    \node (a) at (0,0){\begin{tikzpicture}
        \node[gauge, label=below:$1$] (1tres) at (0,0){};
        \node[gauge, label=below:$2$] (2tres) at (1,0){};
        \node[gauge, label=below:$3$] (3res) at (2,0){};
        \node[gauge, label=below:$2$] (2rres) at (3,0){};
        \node[gauge, label=below:$1$] (1rres) at (4,0){};
        \draw[-] (1tres)--(2tres)--(3res)--(2rres)--(1rres);
        \end{tikzpicture}};
    \node (b) at (0,-3){\begin{tikzpicture}
        \node[gauge, label=below:$1$] (1tres) at (0,0){};
        \node[gauge, label=below:$2$] (2tres) at (1,0){};
        \node[gauge, label=below:$3$] (3res) at (2,0){};
        \node[gauge, label=below:$2$] (2rres) at (3,0){};
        \node[gauge, label=below:$1$] (1rres) at (4,0){};
        \draw[-] (1tres)--(2tres)(2rres)--(1rres);
        \draw[transform canvas={yshift=1.3pt}](2tres)--(3res);
        \draw[transform canvas={yshift=-1.3pt}](2tres)--(3res);
        \draw[-] (1.6,-0.2)--(1.4,0)--(1.6,0.2);
        \draw[transform canvas={yshift=1.3pt}](2rres)--(3res);
        \draw[transform canvas={yshift=-1.3pt}](2rres)--(3res);
        \draw[-] (2.4,-0.2)--(2.6,0)--(2.4,0.2);
        \end{tikzpicture}};
        \draw[->] (a)--(b) node[midway, right]{\textrm{Lacing}};
        \end{tikzpicture}
    \caption{}
    \label{fig:E6SO3Lace}
    \end{subfigure}
    \caption{$\sorm(3)$ \hyperref[fig:SO2np1QQS]{quotient quiver subtraction} on the affine $E^{(1)}_6$ quiver which occurs in two steps. The first step is shown in \Figref{fig:E6SO3Sub} and involves the subtraction of the $\sorm(3)$ quotient quiver from one maximal leg. The second step, shown in \Figref{fig:E6SO3Lace}, is the non-simple lacing of the two edges going from the $\urm(3)$ node producing  \Quiver{fig:E6SO3QQS}.}
    \label{fig:E6SO3QQS}
\end{figure}
The same moduli space may be found through an explicit computation with Weyl integration. This requires the following embedding of $E_6\hookleftarrow A_2\times A_2\times A_2\hookleftarrow A_2\times A_2\times A_1$ which decomposes the $[1,0,0,0,0,0]$ of $E_6$ as \begin{equation}
    \left(\mu_1\right)_{E_6}\rightarrow \mu_2 \nu_2 + (\mu_1 + \nu_1) \rho^2
\end{equation} where the $\mu, \nu,$ and $\rho$ are highest weight fugacities for $A_2,\;A_2,\;$and $A_1$ respectively. The integration is performed with respect to the $A_1$ fugacity.

This is the second hyper-Kähler quotient construction of the Coulomb branch of \Quiver{fig:E6SO3QQS} using exceptional nilpotent orbits. The first used $\surm(3)$ chain polymerisation of two affine $F_4^{(1)}$ quivers \cite{Hanany:2024fqf, Bennett:2024llh} and yields the following equalities \begin{equation}
    \overline{min. E_6}///\sorm(3)=\left(\overline{min. F_4}\times\overline{min. F_4}\right)///\surm(3)=\mathcal C\left[\text{\Quiver{fig:E6SO3QQS}}\right]
\end{equation}

\subsection{$\overline{min. F_4}///\sorm(3)$}
The affine $F_4^{(1)}$ quiver's $(1)-(2)-(3)-$ long leg is also amenable to an $\sorm(3)$ \hyperref[fig:SO2np1QQS]{quotient quiver subtraction}. This operation is shown in \Figref{fig:F4SO3QQS} and produces \Quiver{fig:F4SO3QQS}, with the subtraction of the quotient quiver in \Figref{fig:F4SO3Sub} and the non-simple lacing in \Figref{fig:F4SO3Lace}. It is important in the non-simple lacing step to note that the edge connecting the $\urm(3)$ and $\urm(2)$ goes from multiplicity two to multiplicity four.

The refined Coulomb branch Hilbert series of \Figref{fig:F4SO3QQS} is cumbersome to present, however the unrefined Coulomb branch Hilbert series is presented as \begin{equation}
    \hsC{fig:F4SO3QQS}=\frac{1 + 3 t^2 + 31 t^4 + 55 t^6 + 156 t^8 + 132 t^{10} + 156 t^{12} + 
 55 t^{14} + 31 t^{16} + 3 t^{18} + t^{20}}{(1 - t^2)^5 (1 - t^4)^5}
\end{equation}This moduli space has no particular name as a symplectic singularity but it does have $\surm(3)$ symmetry which can be confirmed from the refined Hilbert series (not presented here). Further support for this global symmetry takes of the form of the Coulomb branch Hasse diagram drawn in \Figref{fig:F4SO3CoulHasse}.
\begin{figure}[h!]
    \centering
    \begin{subfigure}{0.4\textwidth}
    \begin{tikzpicture}
       \node[gauge, label=below:$1$] (1) at (0,0){};
        \node[gauge, label=below:$2$] (2) at (1,0){};
        \node[gauge, label=below:$3$] (3) at (2,0){};
        \node[gauge, label=below:$2$] (2r) at (3,0){};
        \node[gauge, label=below:$1$] (1r) at (4,0){};
        \draw[-] (1)--(2)--(3) (2r)--(1r);
        \draw[transform canvas={yshift=1.3pt}](3)--(2r);
        \draw[transform canvas={yshift=-1.3pt}](3)--(2r);
        \draw[-] (2.4,0.2)--(2.6,0)--(2.4,-0.2);
        \node[gauge, label=below:$1$] (1sub) at (0,-1){};
        \node[gauge, label=below:$2$] (2sub) at (1,-1){};
        \draw[-] (1sub)--(2sub);
        \node[] (minus) at (-1,-1) {$-$};
        \node[gauge, label=below:$3$] (3res) at (2,-2){};
        \node[gauge, label=below:$2$] (2rres) at (3,-2){};
        \node[gauge, label=below:$1$] (1rres) at (4,-2){};
        \draw[-] (2rres)--(1rres);
        \draw[transform canvas={yshift=1.3pt}](3res)--(2rres);
        \draw[transform canvas={yshift=-1.3pt}](3res)--(2rres);
        \draw[-] (2.4,-1.8)--(2.6,-2)--(2.4,-2.2);
    \end{tikzpicture}     
    \caption{}
    \label{fig:F4SO3Sub}
    \end{subfigure}
    \begin{subfigure}{0.4\textwidth}
    \centering
        \begin{tikzpicture}
        \node (a) at (0,0){\begin{tikzpicture}
        \node[gauge, label=below:$3$] (3res) at (2,-2){};
        \node[gauge, label=below:$2$] (2rres) at (3,-2){};
        \node[gauge, label=below:$1$] (1rres) at (4,-2){};
        \draw[-] (2rres)--(1rres);
        \draw[transform canvas={yshift=1.3pt}](3res)--(2rres);
        \draw[transform canvas={yshift=-1.3pt}](3res)--(2rres);
        \draw[-] (2.4,-1.8)--(2.6,-2)--(2.4,-2.2);
        \end{tikzpicture}};
    \node (b) at (0,-3){\begin{tikzpicture}
        \node[gauge, label=below:$3$] (3res) at (2,-2){};
        \node[gauge, label=below:$2$] (2rres) at (3,-2){};
        \node[gauge, label=below:$1$] (1rres) at (4,-2){};
        \draw[-] (2rres)--(1rres);
        \draw[transform canvas={yshift=3pt}](3res)--(2rres);
        \draw[transform canvas={yshift=1pt}](3res)--(2rres);
        \draw[transform canvas={yshift=-1pt}](3res)--(2rres);
        \draw[transform canvas={yshift=-3pt}](3res)--(2rres);
        \draw[-] (2.4,-1.8)--(2.6,-2)--(2.4,-2.2);
        \end{tikzpicture}};
        \draw[->] (a)--(b) node[midway, right]{\textrm{Lacing}};
        \end{tikzpicture}
    \caption{}
    \label{fig:F4SO3Lace}
    \end{subfigure}
    \begin{subfigure}{0.18\textwidth}
    \centering
    \begin{tikzpicture}
    \node[hasse] (1) at (0,0){};
    \node[hasse] (2) at (0,-1.5){};
    \node[hasse] (3) at (0,-3){};
    \draw[-] (1)--(2)node[midway, left]{$\mathcal Y(4)$}--(3) node[midway, left]{$h_{3,4}$};
    \end{tikzpicture}     
    \caption{}
    \label{fig:F4SO3CoulHasse}
    \end{subfigure}
    \caption{$\sorm(3)$ \hyperref[fig:SO2np1QQS]{quotient quiver subtraction} on the affine $F^{(1)}_4$ quiver occurs in two steps. The first step is shown in \Figref{fig:F4SO3Sub} and involves the subtraction of the $\sorm(3)$ quotient quiver from one maximal leg. The second step, shown in \Figref{fig:F4SO3Lace}, is the non-simple lacing of the two edges going from the $\urm(3)$ node producing  \Quiver{fig:F4SO3QQS}. Note that the non-simply laced-ness of this edge goes from two to four. The Coulomb branch Hasse diagram is given in \Figref{fig:F4SO3CoulHasse}.}
    \label{fig:F4SO3QQS}
\end{figure}
The same moduli space emerges from Weyl integration with the embedding $F_4\hookleftarrow A_2\times A_2\hookleftarrow A_2\times A_1$ decomposing the $[1,0,0,0]$ of $F_4$ as \begin{equation}
    \left(\mu_1\right)_{F_4}\rightarrow \mu_1\mu_2 +\left(1+\mu_1^2+\mu_2^2\right)\nu^2+\nu^4
\end{equation} Note that $\mu$ and $\nu$ are highest weight fugacities for $A_2$ and $A_1$ respectively. The integration is performed with respect to the $A_1$ fugacity.

This quiver \Quiver{fig:F4SO3QQS} may be thought of as the folding \cite{Bourget:2020bxh} of \Quiver{fig:E6SO3QQS}. This is an example of the commutation of quotient quiver subtraction and discrete operations on quivers, first noted in \cite{Hanany:2023tvn}.

The quiver \Quiver{fig:F4SO3QQS} is also believed to be a magnetic quiver for the $4d\;\mathcal N=2$ rank-1 theory with flavour symmetry $A_2$. This theory has been conjectured to be derived from the $T^2$ compactification with $\mathbb Z_4$ twist of the $6d\;\mathcal N=(1,0)\;\surm(4)$ gauge theory with 12 flavours and an anti-symmetric \cite{Ohmori:2018ona}. The $\sorm(3)$ quotient quiver subtraction in \Figref{fig:F4SO3QQS} is suggestive of another construction of the Higgs branch of the rank-1 $A_2$ theory \begin{equation}
    \mathcal H(T_3^{\mathbb Z_2})///\sorm(3)=\overline{min. F_4}///\sorm(3)=\mathcal H(A_2)
\end{equation}where the $T^3$ theory enjoys a $\mathbb Z_2$ twist which changes the Higgs branch from $\overline{min. E_6}$ to $\overline{min. F_4}$ \cite{1992math......4227B}, seen in the magnetic quiver through folding \cite{Bourget:2020bxh}.
\section{\boldmath$\sprm(n)+\frac{1}{2}\mathsf{F}\;$\unboldmath Quotient Quivers}
\label{sec:Sp1FQQS}
\begin{figure}
    \centering
    \begin{subfigure}[t]{0.4\textwidth}
    \centering
        \begin{tikzpicture}
        \node[gauge, label=below:$N$] (n)at (0,0){};
        \node[gaugeb, label=below:$\sprm(n)$] (spr) at (-1,0){};
        \node[flavour, label=above:$2N+k-2n$] (flav) at (0,1){};
        \node[flavourr, label=left:$\orm(1)$] (flav2) at (-1,1){};
        \draw[-] (flav2)--(spr)--(n)--(flav);
    \end{tikzpicture}
    \caption{}
    \label{fig:UnSpHtheory}
    \end{subfigure}
    \begin{subfigure}[t]{0.4\textwidth}
    \centering
        \begin{tikzpicture}
        \draw[-] (0,-1)--(0,1) (-2,-1)--(-2,1);
        \node[O5circle,label=below:$\widetilde{\mathrm{O5}^-}$] at (-4,0){};

        \draw[-] (0,0)--(-2,0)node[midway, above]{$N$}--(-3.8,0)node[midway, above]{$2n$};
        \node[D5] at (-0.5,0.75){};
        \node[D5] at (-1.5,0.75){};
        \node at (-1,0.75){$\cdots$};
        \draw [decorate, decoration = {brace, raise=10pt, 
         amplitude=5pt}] (-1.5,0.75) --  (-0.5,0.75) node[pos=0.5,above=15pt,black]{$2N+k-2n$};  
    \end{tikzpicture}
    \caption{}
    \label{fig:UnSpHBrane}
    \end{subfigure}
    \begin{subfigure}[t]{0.75\textwidth}
    \begin{tikzpicture}
         \node[gauge, label=below:$1$] (1l) at (0,0){};
        \node[gauge, label=below:$2$] (2l) at (-1,0){};
        \node[] (cdotsl) at (-2,0){$\cdots$};
        \node[gauge, label=below:$N$] (nl) at (-3,0){};
        \node[gauge, label=below:$N$] (nml) at (-4,0){};
        \node[] (cdotsm) at (-5,0){$\cdots$};
        \node[gauge, label=below:$N$] (nmr) at (-6,0){};
        \node[gauge, label=below:$N$] (nr) at (-7,0){};
        \node[gauge, label=below:$N-1$] (nm1) at (-8,0){};
        \node[] (cdotsr) at (-9,0){$\cdots$};
        \node[gauge, label=below:$2n+1$] (2rp3) at (-10,0){};
        \node[gauge, label=below:$n$] (2rp2) at (-11,0){};
        
        \node[flavour, label=above:$1$] (1fl) at (-3,1){};
        \node[flavour, label=above:$1$] (1fr) at (-7,1){};

        \draw[-] (1l)--(2l)--(cdotsl)--(nl)--(nml)--(cdotsm)--(nmr)--(nr)--(nm1)--(cdotsr)--(2rp3) (1fl)--(nl) (1fr)--(nr);

        \draw[transform canvas={yshift=1.3pt}] (2rp3)--(2rp2);
        \draw[transform canvas={yshift=-1.3pt}] (2rp3)--(2rp2);

        \draw[-] (-10.4,0.2)--(-10.6,0)--(-10.4,-0.2);

        \draw [decorate, decoration = {brace, raise=15pt, 
         amplitude=5pt}] (-3,0) --  (-7,0) node[pos=0.5,below=20pt,black]{$k+1$};
    \end{tikzpicture}
    \caption{}
    \label{fig:UnSpHtheoryMirror}
    \end{subfigure}
    \begin{subfigure}[t]{0.75\textwidth}
    \centering
    \begin{tikzpicture}
        \draw[-] (0,-1)--(0,1) (-1,-1)--(-1,1) (-2,-1)--(-2,1) (-3,-1)--(-3,1) (-4,-1)--(-4,1) (-5,-1)--(-5,1) (-6,-1)--(-6,1) (-7,-1)--(-7,1) (-8,-1)--(-8,1) (-9,-1)--(-9,1) (-10,-1)--(-10,1) (-11,-1)--(-11,1) (-12,-1)--(-12,1);
        \draw[dashed] (-13,-1)--(-13,1)node[pos=1,above]{$\widetilde{\mathrm{ON}^-}$};
        \draw[-] (0,0)--(-1,0)node[midway, above]{$1$}--(-2,0)node[midway, above]{$2$};
        \draw[-] (-2,0)--(-2.2,0) (-2.8,0)--(-3,0);
        \node at (-2.5,0){$\cdots$};
        \draw[-] (-3,0)--(-4,0)node[midway, above]{$N$};
        \node[D5] at (-3.5,0.75){};
        \draw[-] (-4,0)--(-5,0)node[midway, above]{$N$};
        \draw[-] (-5,0)--(-5.2,0) (-5.8,0)--(-6,0);
        \node at (-5.5,0){$\cdots$};
        \draw[-] (-6,0)--(-7,0)node[midway,above]{$N$}--(-8,0)node[midway,above]{$N$}--(-9,0)node[midway, above]{$N-1$};
        \node[D5] at (-7.5,0.75){};
        \draw[-] (-9,0)--(-9.2,0) (-9.8,0)--(-10,0);
        \node at (-9.5,0){$\cdots$};
        \draw[-] (-10,0)--(-11,0)node[midway, above]{$2n+2$}--(-12,0)node[midway,above]{$2n+1$}--(-13,0)node[midway, above]{$2n$};
        \draw [decorate, decoration = {brace, raise=5pt, 
         amplitude=5pt}] (-8,1) --  (-3,1) node[pos=0.5,above=10pt,black]{$k+2$};
    \end{tikzpicture}
    \caption{}
    \label{fig:UnSpHMirrorBrane}
\end{subfigure}
    \caption{Gauging an $\sprm(n)+\mathsf{\frac{1}{2}}$ flavour subgroup of $\urm(N)$ SQCD with $2N+k$ flavours returns the theory in \Figref{fig:UnSpHtheory}, realised from the brane system in \Figref{fig:UnSpHBrane}. The magnetic theory is conjectured to be \Figref{fig:UnSpHtheoryMirror}, associated to the brane system in \Figref{fig:UnSpHMirrorBrane}, related to that of \Figref{fig:UnSpHBrane} by brane-creation and S-duality. Comparison to the mirror of the SQCD in \eqref{quiv:UN_SQCD} leads to the conjectural quotient quiver \eqref{spn+1/2F_quotient_quiver}.}
    \label{}
\end{figure}

In Sections \ref{sec:Sp(n)QQS}, \ref{sec:So(2n)QQS}, and \ref{sec:So(2n+1)QQS} a quotient quiver subtraction algorithm was presented to gauge $\sprm(n)$ and $\sorm(n)$ subgroups of Coulomb branch global symmetries. Key to these derivations were Type IIB brane systems in the electric phase in the presence of $\mathrm{O5}^{\mp}$ and $\widetilde{\mathrm{O}5^+}$.

The final orientifold to complete the quartet is the $\widetilde{\mathrm{O}5^-}$. The starting point for the analysis is the $\urm(N)$ SQCD with $2N+k$ flavours where $k$ is positive, this parameterisation is for convenience but will not play a role in the resulting quotient quiver subtraction algorithm. Now consider coupling a half-hypermultiplet of $\sprm(n)$ and gauge the $\sprm(n)$ resulting in 
\begin{equation}
    \begin{tikzpicture}
        \node[gauge, label=below:$N$] (n)at (0,0){};
        \node[gaugeb, label=below:$\sprm(n)$] (spr) at (1,0){};
        \node[flavour, label=above:$2N+k-2n$] (flav) at (0,1){};
        \node[flavourr, label=right:$\orm(1)$] (flav2) at (1,1){};
        \draw[-] (flav2)--(spr)--(n)--(flav);
    \end{tikzpicture}\label{eq:UnSp1Ftheory}
\end{equation}

Immediately it is clear that this theory suffers from a Witten anomaly \cite{Witten:1982fp} due to the half-hypermultiplet coupled to $\sprm(n)$. As an academic exercise, one may consider the effect of this operation of gauging $\sprm(n)$ with a half-hyper on the magnetic quiver with the aim of this gauging to construct moduli spaces.

Comparing the magnetic theory of the SQCD before \eqref{quiv:UN_SQCD} and after gauging (\Figref{fig:UnSpHtheoryMirror}) identifies the quotient quiver \eqref{spn+1/2F_quotient_quiver} and the subtraction prescription. The $\sprm(n)+\mathsf{\frac{1}{2}F}$ quotient quiver is
\begin{equation}
    \begin{tikzpicture}
        \node[gauge, label=below:$1$] (1l) at (0,0){};
        \node[gauge, label=below:$2$] (2l) at (1,0){};
        \node[] (cdots) at (2,0){$\cdots$};
        \node[gauge, label=above:$2n-2$] (2rm1) at (3,0){};
        \node[gauge, label=below:$2n-1$] (2r) at (4,0){};
        \draw[-] (1l)--(2l)--(cdots)--(2rm1)--(2r);
    \end{tikzpicture}
    \label{spn+1/2F_quotient_quiver}
\end{equation}which has total gauge node rank of $n(2n-1)$, which is $n$ short of the expected dimension reduction of gauging $\sprm(n)$ and adding a half-hyper, $2n^2$. However one should note that the leftmost gauge node in the quiver shown in \Figref{fig:UnSpHtheoryMirror} is $\urm(n)$ rather than $\urm(2n)$, this reduction of the rank by $n$ compensates for the total rank of gauge nodes in the quotient quiver.

\paragraph{Formal Statement of the Rule}
\begin{enumerate}
    \item Take a target quiver with a long leg of gauge nodes up to at least $\urm(2n)$. The edge in \textcolor{teal}{teal} denotes some generic connection to the rest of the quiver $\mathcal Q$ -- the only restriction is that the long leg must correspond to long roots of the Coulomb branch global symmetry algebra.
    \begin{equation}
    \begin{tikzpicture}
    \node[gauge, label=below:$1$] (1) at (0,0){};
    \node[gauge, label=below:$2$] (2) at (1,0){};
    \node[] (cdots) (cdots) at (2,0){$\cdots$};
    \node[gauge, label=below:$2n-1$] (2r) at (3,0){};
    \node[gauge, label=below:$2n$] (2rp1) at (4,0){};
    \node[gauge] (Q) at (5,0){$Q$};
    \draw[-] (1)--(2)--(cdots)--(2r)--(2rp1);
    \draw[teal,very thick] (2rp1)--(Q);   
    \end{tikzpicture}  
    \end{equation}
    \item Align end of the $\sprm(n)+\frac{1}{2}\mathsf{F}$ quotient quiver against the leg of the target quiver and subtract the ranks of the quotient quiver from those of the target. This deletes the gauge nodes from $\urm(1)$ up to $\urm(2n-1)$ in the maximal chain, leaving only $\urm(2n)$. Rebalancing is not required.
    \begin{equation}
    \begin{tikzpicture}
    \node[gauge, label=below:$1$] (1) at (0,0){};
    \node[gauge, label=below:$2$] (2) at (1,0){};
    \node[] (cdots) (cdots) at (2,0){$\cdots$};
    \node[gauge, label=below:$2n-1$] (2r) at (3,0){};
    \node[gauge, label=below:$2n$] (2rp1) at (4,0){};
    \node[gauge] (Q) at (5,0){$Q$};
    \draw[-] (1)--(2)--(cdots)--(2r)--(2rp1);
    \draw[teal,very thick] (2rp1)--(Q);
    \node[gauge, label=below:$1$] (1sub) at (0,-1){};
    \node[gauge, label=below:$2$] (2sub) at (1,-1){};
    \node[] (cdots) (cdotssub) at (2,-1){$\cdots$};
    \node[gauge, label=below:$2n-1$] (2rsub) at (3,-1){};
    \node[] (minus) at (-1,-1) {$-$};
    \draw[-] (1sub)--(2sub)--(cdotssub)--(2rsub);
    \node[gauge, label=below:$2n$] (2rp1r) at (4,-2){};
    \node[gauge] (Qr) at (5,-2){$Q$};
    \draw[teal,very thick] (2rp1r)--(Qr);      
    \end{tikzpicture}  
    \end{equation}
    \item The resulting $\urm(2n)$ gauge node has its rank halved to become $\urm(n)$ and all edges in \textcolor{teal}{teal} double their non-simply lacedness with the $\urm(n)$ node being short.
    \begin{equation}
    \begin{tikzpicture}
    \node(c) at (-5,0){\begin{tikzpicture}
         \node[gauge, label=below:$2n$] (2rp1) at (0,0){};
    \node[gauge] (Qr) at (1,0){$Q$};
    \draw[teal,very thick] (2rp1)--(Qr); 
    \end{tikzpicture}};
    \node (a) at (0,0){\begin{tikzpicture}
    \node[gauge, label=below:$n$] (2rp1) at (0,0){};
    \node[gauge] (Qr) at (1,0){$Q$};
    \draw[teal,very thick] (2rp1)--(Qr); 
    \end{tikzpicture}};
    \node (b) at (5,0){\begin{tikzpicture}
    \node[gauge, label=below:$n$] (r) at (0,0){};
    \node[gauge] (Q) at (1,0){$Q$};
    \draw[teal, very thick, transform canvas={yshift=1.3pt}] (r)--(Q);
    \draw[teal, very thick, transform canvas={yshift=-1.3pt}] (r)--(Q);
    \draw[teal, very thick, -] (0.6,0.2)--(0.4,0)--(0.6,-0.2);
    \end{tikzpicture}};
    \draw[->] (c)--node[midway, above]{\textrm{Halve rank}}(a);
    \draw[->] (a)--(b) node[midway, above]{\textrm{Double lacing}};
    \end{tikzpicture}
    \end{equation}
\end{enumerate}

\subsection{$\left(\overline{min. E_6}\times\mathbb H\right)///\sprm(1)$}
A simple but non-trivial example of the $\sprm(1)+\frac{1}{2}\mathsf{F}$ quotient quiver subtraction is performed on the affine $E^{(1)}_6$ quiver to produce \Quiver{fig:E6Sp1HQQS} as shown in \Figref{fig:E6Sp1HQQS}.
\begin{figure}[h!]
    \centering
    \begin{subfigure}{0.45\linewidth}
    \centering
    \begin{tikzpicture}
        \node[gauge, label=below:$1$] (1l) at (0,0){};
        \node[gauge, label=below:$2$] (2l) at (1,0){};
        \node[gauge, label=below:$3$] (3) at (2,0){};
        \node[gauge, label=below:$2$] (2r) at (3,0){};
        \node[gauge, label=below:$1$] (1r) at (4,0){};
        \node[gauge, label=right:$2$] (2t) at (2,1){};
        \node[gauge, label=above:$1$] (1t) at (2,2){};
        \draw[-] (1l)--(2l)--(3)--(2r)--(1r) (3)--(2t)--(1t);
        \node[gauge, label=below:$1$] (1sub) at (0,-1){};
        \node[] (minus) at (-1,-1) {$-$};

        \node[gauge, label=above:$1$] (1tres) at (2,-2){};
        \node[gauge, label=right:$2$] (2tres) at (2,-3){};
        \node[gauge, label=below:$2$] (2lres) at (1,-4){};
        \node[gauge, label=below:$3$] (3res) at (2,-4){};
        \node[gauge, label=below:$2$] (2rres) at (3,-4){};
        \node[gauge, label=below:$1$] (1rres) at (4,-4){};
        \draw[-] (2lres)--(3res) (1tres)--(2tres)--(3res)--(2rres)--(1rres);
    \end{tikzpicture}     
    \caption{}
    \label{fig:E6Sp1HSub}
    \end{subfigure}
    \begin{subfigure}{0.45\linewidth}
    \centering
    \begin{tikzpicture}
    \node (a) at (0,0){\begin{tikzpicture}
         \node[gauge, label=above:$1$] (1tres) at (2,-2){};
        \node[gauge, label=right:$2$] (2tres) at (2,-3){};
        \node[gauge, label=below:$2$] (2lres) at (1,-4){};
        \node[gauge, label=below:$3$] (3res) at (2,-4){};
        \node[gauge, label=below:$2$] (2rres) at (3,-4){};
        \node[gauge, label=below:$1$] (1rres) at (4,-4){};
        \draw[-] (2lres)--(3res) (1tres)--(2tres)--(3res)--(2rres)--(1rres);
        \end{tikzpicture}};
    \node (b) at (0,-4){\begin{tikzpicture}
         \node[gauge, label=above:$1$] (1tres) at (2,-2){};
        \node[gauge, label=right:$2$] (2tres) at (2,-3){};
        \node[gauge, label=below:$1$] (1lres) at (1,-4){};
        \node[gauge, label=below:$3$] (3res) at (2,-4){};
        \node[gauge, label=below:$2$] (2rres) at (3,-4){};
        \node[gauge, label=below:$1$] (1rres) at (4,-4){};
        \draw[-] (1tres)--(2tres)--(3res)--(2rres)--(1rres);
        \draw[transform canvas={yshift=1.3pt}](1lres)--(3res);
        \draw[transform canvas={yshift=-1.3pt}](1lres)--(3res);
        \draw[-] (1.6,-4.2)--(1.4,-4)--(1.6,-3.8);
        \end{tikzpicture}};
        \draw[->] (a)--(b) node[midway, right]{\textrm{Lacing}};
        \end{tikzpicture}
    \caption{}
    \label{fig:E6Sp1HLace}
    \end{subfigure}
    
    \caption{$\sprm(1)+\frac{1}{2}\mathsf{F}$ quotient quiver subtraction on the affine $E^{(1)}_6$ quiver which occurs in two steps. The first step is shown in \Figref{fig:E6Sp1HSub} and involves the subtraction of the $\sprm(1)+\frac{1}{2}\mathsf{F}$ quotient quiver from one maximal leg. The second step, shown in \Figref{fig:E6Sp1HLace}, is the non-simple lacing of the $\urm(2)$ gauge node producing \Quiver{fig:E6Sp1HQQS}.}
    \label{fig:E6Sp1HQQS}
\end{figure}

The Coulomb branch Hilbert series may be computed exactly, but for brevity the unrefined Hilbert series is presented here as \begin{equation}
    \hsC{fig:E6Sp1HQQS}=\frac{\left(\begin{aligned}1 &+ 17 t^2 + 20 t^3 + 117 t^4 + 180 t^5 + 392 t^6 + 592 t^7 + 
 701 t^8 + 872 t^9 \\&+ 701 t^{10} + 592 t^{11} + 392 t^{12} + 180 t^{13} + 
 117 t^{14} + 20 t^{15} + 17 t^{16} + t^{18}\end{aligned}\right)}{(1 - t^2)^{18}}
\end{equation}which does not elucidate the identity of this moduli space. However the Highest Weight Generating function (HWG), computed from the refined Hilbert series is given as \begin{equation}
    \hwg\left[\mathcal C\left(\text{\Quiver{fig:E6Sp1HQQS}}\right)\right]=\pe\left[\mu_1\mu_5t^2+\mu_3t^3+\mu_2\mu_4t^4\right]
\end{equation}identifies \begin{equation}
    \mathcal C\left(\text{\Quiver{fig:E6Sp1HQQS}}\right)=\mathbb Z_2\;\text{Cover}\;\overline{\mathcal O}^{A_5}_{[0,0,2,0,0]}
\end{equation} The $\mathbb Z_2$ cover is apparent in the magnetic quiver \Quiver{fig:E6Sp1HQQS} since the difference between this quiver and the magnetic quiver for $\overline{\mathcal O}^{A_5}_{[0,0,2,0,0]}$, drawn in \eqref{quiv:A5Orbit}, is the non-simply laced edge of multiplicity two \cite{Hanany:2020jzl}.

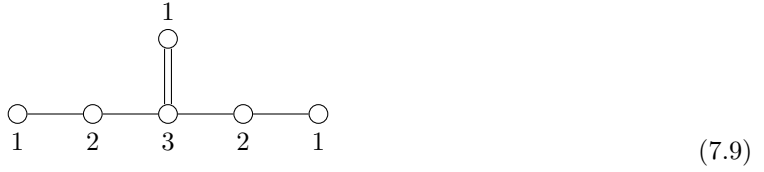
\begin{equation}
    \begin{tikzpicture}
        \node[gauge, label=below:$1$] (1l) at (0,0){};
        \node[gauge, label=below:$2$] (2l) at (1,0){};
        \node[gauge, label=below:$3$] (3) at (2,0){};
        \node[gauge, label=below:$2$] (2r) at (3,0){};
        \node[gauge, label=below:$1$] (1r) at (4,0){};
        \node[gauge, label=above:$1$] (1t) at (2,1){};

        \draw[-] (1l)--(2l)--(3)--(2r)--(1r);
        \draw[transform canvas={xshift=1.3pt}](3)--(1t);
        \draw[transform canvas={xshift=-1.3pt}](3)--(1t);
        
    \end{tikzpicture}
    \label{quiv:A5Orbit}
\end{equation}

This result is verified with Weyl integration with the embedding of $\sprm(1)$ inside $E_6$ with commutant $\surm(6)$ decomposing the adjoint of $E_6$ as \begin{equation}
    \left(\mu_6\right)_{E_6}\rightarrow \mu_1\mu_5+\nu^2+\mu_3\nu
\end{equation}where the $\mu_i$ refer to $\surm(6)$ highest weights fugacity and $\nu$ refers to the $\sprm(1)$ highest weight fugacity. If an additional fugacity, $a$, for the $\orm(1)$ is included, the HWG of the hyper-Kähler quotient is given as \begin{equation}
     \hwg\left[\left(\overline{min. E_6}\times\mathbb H\right)///\sprm(1)\right]=\pe\left[\mu_1\mu_5t^2+a\mu_3t^3+\mu_2\mu_4t^4\right]
\end{equation}which explicitly gives the construction \begin{equation}
     \left(\overline{min. E_6}\times\mathbb H\right)///\sprm(1)=\mathcal C\left(\text{\Quiver{fig:E6Sp1HQQS}}\right)=\mathbb Z_2\;\text{Cover}\;\overline{\mathcal O}^{A_5}_{[0,0,2,0,0]}
\end{equation} and as discrete quotients and hyper-Kähler quotients tend to commute, a simple corollary by gauging the $\mathbb Z_2$ in $\mathbb H$ \cite{Benvenuti:2010pq} \begin{equation}
    \left(\overline{min. E_6}\times\mathbb C^2/\mathbb Z_2\right)///\sprm(1)=\overline{\mathcal O}^{A_5}_{[0,0,2,0,0]}
\end{equation} which is related to the duality between the $\sprm(1)$ gauging of the rank-1 $E_6$ theory coupled to two free hypermultiplets and $\surm(3)$ with six flavours \cite{Argyres:2007tq, Hanany:2023tvn}.
\subsection{$\left(\overline{min. E_7}\times\mathbb H^2\right)///\sprm(2)$}
An example of the $\sprm(2)+\frac{1}{2}\mathsf{F}$ quotient quiver subtraction is performed on the affine $E^{(1)}_7$ quiver to produce \Quiver{fig:E7Sp2HQQS} as shown in \Figref{fig:E7Sp2HQQS}.
\begin{figure}[h!]
    \centering
    \begin{subfigure}{0.45\linewidth}
    \centering
    \begin{tikzpicture}
       \node[gauge, label=below:$1$] (1l) at (0,0){};
        \node[gauge, label=below:$2$] (2l) at (1,0){};
        \node[gauge, label=below:$3$] (3l) at (2,0){};
        \node[gauge, label=below:$4$] (4) at (3,0){};
        \node[gauge, label=below:$3$] (3r) at (4,0){};
        \node[gauge, label=below:$2$] (2r) at (5,0){};
        \node[gauge, label=below:$1$] (1r) at (6,0){};
        \node[gauge, label=above:$2$] (2t) at (3,1){};
        \draw[-] (1l)--(2l)--(3l)--(4)--(3r)--(2r)--(1r) (2t)--(4);
        \node[gauge, label=below:$1$] (1sub) at (0,-1){};
        \node[gauge, label=below:$2$] (2sub) at (1,-1){};
        \node[gauge, label=below:$3$] (3sub) at (2,-1){};
        \node[] (minus) at (-1,-1) {$-$};
        \draw[-] (1sub)--(2sub)--(3sub);
        \node[gauge, label=above:$2$] (2tres) at (3,-2){};
        \node[gauge, label=below:$4$] (4res) at (3,-3){};
        \node[gauge, label=below:$3$] (3res) at (4,-3){};
        \node[gauge, label=below:$2$]  (2res) at (5,-3){};
        \node[gauge, label=below:$1$] (1res) at (6,-3){};
        \draw[-] (2tres)--(4res)--(3res)--(2res)--(1res);
    \end{tikzpicture}     
    \caption{}
    \label{fig:E7Sp2HSub}
    \end{subfigure}
    \begin{subfigure}{0.45\linewidth}
    \centering
    \begin{tikzpicture}
    \node (a) at (0,0){\begin{tikzpicture}
        \node[gauge, label=above:$2$] (2tres) at (3,-2){};
        \node[gauge, label=below:$4$] (4res) at (3,-3){};
        \node[gauge, label=below:$3$] (3res) at (4,-3){};
        \node[gauge, label=below:$2$]  (2res) at (5,-3){};
        \node[gauge, label=below:$1$] (1res) at (6,-3){};
        \draw[-] (2tres)--(4res)--(3res)--(2res)--(1res);
        \end{tikzpicture}};
    \node (b) at (0,-4){\begin{tikzpicture}
        \node[gauge, label=below:$2$] (2tres) at (2,-3){};
        \node[gauge, label=below:$2$] (4res) at (3,-3){};
        \node[gauge, label=below:$3$] (3res) at (4,-3){};
        \node[gauge, label=below:$2$]  (2res) at (5,-3){};
        \node[gauge, label=below:$1$] (1res) at (6,-3){};
        \draw[-] (3res)--(2res)--(1res);

        \draw[transform canvas={yshift=1.3pt}](2tres)--(4res);
        \draw[transform canvas={yshift=-1.3pt}](2tres)--(4res);
        \draw[-] (2.4,-3.2)--(2.6,-3)--(2.4,-2.8);
        \draw[transform canvas={yshift=1.3pt}](4res)--(3res);
        \draw[transform canvas={yshift=-1.3pt}](4res)--(3res);
        \draw[-] (3.6,-3.2)--(3.4,-3)--(3.6,-2.8);
        \end{tikzpicture}};
        \draw[->] (a)--(b) node[midway, right]{\textrm{Lacing}};
        \end{tikzpicture}
    \caption{}
    \label{fig:E7Sp2HLace}
    \end{subfigure}
    
    \caption{$\sprm(2)+\frac{1}{2}\mathsf{F}$ quotient quiver subtraction on the affine $E^{(1)}_7$ quiver which occurs in two steps. The first step is shown in \Figref{fig:E7Sp2HSub} and involves the subtraction of the $\sprm(2)+\frac{1}{2}\mathsf{F}$ quotient quiver from one maximal leg. The second step, shown in \Figref{fig:E7Sp2HLace}, is the non-simple lacing of the $\urm(4)$ gauge node producing \Quiver{fig:E7Sp2HQQS}.}
    \label{fig:E7Sp2HQQS}
\end{figure}

The Coulomb branch Hilbert series may be computed exactly, but for brevity the unrefined Hilbert series is presented here as \begin{equation}
    \hsC{fig:E7Sp2HQQS}=\frac{\left(\begin{aligned}1 &+ 13 t^2 + 16 t^3 + 104 t^4 + 192 t^5 + 589 t^6 + 1216 t^7 + 
 2466 t^8 + 4672 t^9 + 7628 t^{10} \\&+ 12160 t^{11} + 16970 t^{12} + 
 22480 t^{13} + 27362 t^{14} + 30400 t^{15} + 32006 t^{16} + 30400 t^{17} \\&+ 
 27362 t^{18} + 22480 t^{19} + 16970 t^{20} + 12160 t^{21} + 7628 t^{22} + 
 4672 t^{23} + 2466 t^{24} \\&+ 1216 t^{25} + 589 t^{26} + 192 t^{27} + 104 t^{28} + 
 16 t^{29} + 13 t^{30} + t^{32}\end{aligned}\right)}{(1-t^2)^{11}(1-t^4)^7}
\end{equation}which does not elucidate the identity of this moduli space. Although the global symmetry is identified as $\surm(2)\times\sorm(7)$.

This result is verified with Weyl integration with the embedding of $\sprm(2)$ inside $E_7$ with commutant $\surm(2)\times \sorm(7)$ decomposing the adjoint of $E_7$ as \begin{equation}
    \left(\mu_1\right)_{E_7}\rightarrow \mu_2 +\rho^2  + \nu_1^2+ \rho \mu_3 \nu_1  + \mu_1 \nu_2
\end{equation}where the $\mu_i$ refer to $B_3$ highest weight fugacities, $\rho$ refers to the $A_1$ highest weight fugacity, and the $\nu$ refer to $\sprm(2)$ highest weight fugacities.

The conclusion is that \begin{equation}
    \left(\overline{min. E_7}\times\mathbb H^2\right)///\sprm(2)=\mathcal C\left(\text{\Quiver{fig:E7Sp2HQQS}}\right)
\end{equation}

\section{Outlook}
\label{sec:outlook}
This paper introduces quotient quiver subtraction prescriptions for the gauging of $\sprm(n)$ (coupled to a half hypermultiplet) and $\sorm(n)$ Coulomb branch global symmetry subgroups for $3d\;\mathcal N=4$ quivers with unitary gauge nodes. This completes the family of prescriptions to gauging classical Coulomb branch global symmetry subgroups in unitary quivers \cite{Hanany:2023tvn, Hanany:2024fqf, Dancer:2024lra}.

There are clear future directions to explore. The first natural extension is to use Type IIB brane systems with $\mathrm{D}3,\;\mathrm{D}5$, and $\mathrm{NS}5$ with $\mathrm{ON}$ planes. This would provide an algorithm on the magnetic quivers to gauge products of unitary groups in the case of $\mathrm{ON}^-$, as well as give an interpretation of non-simply laced edges for the other types of $\mathrm{ON}$. Secondly, it would be possible to include $\mathrm{O}3,\;\mathrm{O}5,$ and $\mathrm{ON}$ which would allow for the exploration of gauging product symmetries or gauging with non-simple lacing of orthosymplectic quivers. This will all be investigated in an upcoming work.

Similarly, a natural extension is to find quotient quiver subtraction algorithms for exceptional groups. To date only $G_2$ gaugings of the Coulomb branch global symmetry have been constructed in unframed orthosymplectic theories \cite{Bennett:2024llh}. It is possible that gauging $G_2$ Coulomb branch global symmetry subgroups may admit a simple combinatorial operation as it lies in between $\sorm(7)$ and $\surm(3)$ in terms of Higgsing patterns for six dimensional theories.
\acknowledgments
The work of SB, AH, and GK is partially supported by STFC Consolidated Grant ST/X000575/1. The work of SB is supported by the STFC DTP research studentship grant ST/Y509231/1. The work of GK is supported by STFC DTP research studentship grant ST/X508433/1.
\appendix
\bibliographystyle{JHEP}
\bibliography{references.bib}
\end{document}